\documentclass[prl,letter,twocolumn,superscriptaddress,floatfix,nofootinbib,showpacs,amsmath,amssymb,altaffilletter,floatfix]{revtex4-1}
\input{ds.def}
\definecolor{ocre}{RGB}{243,102,25}
\definecolor{citrine}{RGB}{228, 208, 10}

\newcommand{\APC}{APC, Universit\'e Paris Diderot, CNRS/IN2P3, CEA/Irfu, Obs de Paris, USPC, Paris 75205, France}
\newcommand{\AQLNGS}{INFN Laboratori Nazionali del Gran Sasso, Assergi (AQ) 67100, Italy}
\newcommand{\AQGSSI}{Gran Sasso Science Institute, L'Aquila 67100, Italy}

\newcommand{\Augustana}{Physics Department, Augustana University, Sioux Falls, SD 57197, USA}
\newcommand{\Belgorod}{Radiation Physics Laboratory, Belgorod National Research University, Belgorod 308007, Russia}
\newcommand{\BHSU}{School of Natural Sciences, Black Hills State University, Spearfish, SD 57799, USA}

\newcommand{\Bnl}{Brookhaven National Laboratory, Upton, NY 11973, USA}
\newcommand{\BOINFN}{INFN Bologna, Bologna 40126, Italy}
\newcommand{\BOUniPHY}{Physics Department, Universit\`a degli Studi di Bologna, Bologna 40126, Italy}

\newcommand{\CAUniPHY}{Physics Department, Universit\`a degli Studi di Cagliari, Cagliari 09042, Italy}
\newcommand{\CAINFN}{INFN Cagliari, Cagliari 09042, Italy}

\newcommand{\Campinas}{Physics Institute, Universidade Estadual de Campinas, Campinas 13083, Brazil}

\newcommand{\CIEMAT}{CIEMAT, Centro de Investigaciones Energ\'eticas, Medioambientales y Tecnol\'ogicas, Madrid 28040, Spain}

\newcommand{\CTLNS}{INFN Laboratori Nazionali del Sud, Catania 95123, Italy}
\newcommand{\ENUniCEE}{Engineering and Architecture Faculty, Universit\`a di Enna Kore, Enna 94100, Italy}

\newcommand{\Fnal}{Fermi National Accelerator Laboratory, Batavia, IL 60510, USA}

\newcommand{\GEUni}{Physics Department, Universit\`a degli Studi di Genova, Genova 16146, Italy}
\newcommand{\GEINFN}{INFN Genova, Genova 16146, Italy}

\newcommand{\Hawaii}{Department of Physics and Astronomy, University of Hawai'i, Honolulu, HI 96822, USA}
\newcommand{\Houston}{Department of Physics, University of Houston, Houston, TX 77204, USA}
\newcommand{\IHEP}{Institute of High Energy Physics, Beijing 100049, China}

\newcommand{\INSTM}{Interuniversity Consortium for Science and Technology of Materials, Firenze 50121, Italy}

\newcommand{\JINR}{Joint Institute for Nuclear Research, Dubna 141980, Russia}
\newcommand{\Krakow}{M. Smoluchowski Institute of Physics, Jagiellonian University, 30-348 Krakow, Poland}
\newcommand{\Kurchatov}{National Research Centre Kurchatov Institute, Moscow 123182, Russia}

\newcommand{\Lodz}{Institute of Applied Radiation Chemistry, Lodz University of Technology, 93-590 Lodz, Poland}
\newcommand{\LPNHE}{LPNHE, CNRS/IN2P3, Sorbonne Universit\'e, Universit\'e Paris Diderot, Paris 75252, France}

\newcommand{\MEPhI}{National Research Nuclear University MEPhI, Moscow 115409, Russia}

\newcommand{\MIINFN}{INFN Milano, Milano 20133, Italy}

\newcommand{\MIUni}{Physics Department, Universit\`a degli Studi di Milano, Milano 20133, Italy}
\newcommand{\MSU}{Skobeltsyn Institute of Nuclear Physics, Lomonosov Moscow State University, Moscow 119234, Russia}
\newcommand{\NAINFN}{INFN Napoli, Napoli 80126, Italy}
\newcommand{\NAUniPHY}{Physics Department, Universit\`a degli Studi ``Federico II'' di Napoli, Napoli 80126, Italy}

\newcommand{\Petersburg}{Saint Petersburg Nuclear Physics Institute, Gatchina 188350, Russia}
\newcommand{\PGUniCBB}{Chemistry, Biology and Biotechnology Department, Universit\`a degli Studi di Perugia, Perugia 06123, Italy}
\newcommand{\PGINFN}{INFN Perugia, Perugia 06123, Italy}
\newcommand{\PIINFN}{INFN Pisa, Pisa 56127, Italy}
\newcommand{\PIUniPHY}{Physics Department, Universit\`a degli Studi di Pisa, Pisa 56127, Italy}
\newcommand{\PNNL}{Pacific Northwest National Laboratory, Richland, WA 99352, USA}
\newcommand{\Princeton}{Physics Department, Princeton University, Princeton, NJ 08544, USA}

\newcommand{\RMTreINFN}{INFN Roma Tre, Roma 00146, Italy}
\newcommand{\RMTreUni}{Mathematics and Physics Department, Universit\`a degli Studi Roma Tre, Roma 00146, Italy}
\newcommand{\RMUnoINFN}{INFN Sezione di Roma, Roma 00185, Italy}
\newcommand{\RMUnoUni}{Physics Department, Sapienza Universit\`a di Roma, Roma 00185, Italy}

\newcommand{\SSUniCHP}{Chemistry and Pharmacy Department, Universit\`a degli Studi di Sassari, Sassari 07100, Italy}

\newcommand{\Temple}{Physics Department, Temple University, Philadelphia, PA 19122, USA}

\newcommand{\UCDavis}{Department of Physics, University of California, Davis, CA 95616, USA}
\newcommand{\UCLA}{Physics and Astronomy Department, University of California, Los Angeles, CA 90095, USA}
\newcommand{\UMass}{Amherst Center for Fundamental Interactions and Physics Department, University of Massachusetts, Amherst, MA 01003, USA}

\newcommand{\USP}{Instituto de F\'isica, Universidade de S\~ao Paulo, S\~ao Paulo 05508-090, Brazil}
\newcommand{\VTech}{Virginia Tech, Blacksburg, VA 24061, USA}
\begin{document}
\title{Low-Mass Dark Matter Search with the \DSf\ Experiment}
\author{P.~Agnes}\affiliation{\Houston}
\author{I.~F.~M.~Albuquerque}\affiliation{\USP}
\author{T.~Alexander}\affiliation{\PNNL}
\author{A.~K.~Alton}\affiliation{\Augustana}
\author{G.~R.~Araujo}\affiliation{\USP}
\author{D.~M.~Asner}\affiliation{\Bnl}
\author{M.~Ave}\affiliation{\USP}
\author{H.~O.~Back}\affiliation{\PNNL}
\author{B.~Baldin}\altaffiliation{Current address: Raleigh, NC 27613-3133, USA}\affiliation{\Fnal}
\author{G.~Batignani}\affiliation{\PIINFN}\affiliation{\PIUniPHY}
% ...

\collaboration{The \DS\ Collaboration}\noaffiliation
\date{\today}
\begin{abstract}
We present the results of a search for dark matter WIMPs in the mass range below \SI{20}{\GeV\per\square\c} using a target of low-radioactivity argon with a \SI{6786.0}{\kg\day} exposure.  The data were obtained using the \DSf\ apparatus at Laboratori Nazionali del Gran Sasso (\LNGS).  The analysis is based on the ionization signal, for which the \DSf\ time projection chamber is fully efficient at \SI{0.1}{\keVee}.  The observed rate in the detector at \SI{0.5}{\keVee} is about \SI{1.5}{event\per\keVee\per\kg\per\day} and is almost entirely accounted for by known background sources.   We obtain a \SI{90}{\percent} C.L. exclusion limit above \SI{1.8}{\GeV\per\square\c} for the spin-independent cross section of dark matter WIMPs on nucleons, extending the exclusion region for dark matter below previous limits in the range \SIrange[range-phrase=--,range-units=single]{1.8}{6}{\GeV\per\square\c}.
\end{abstract}
\maketitle
The concept of dark matter was developed~\cite{Oort:1932tb,Zwicky:1933ub,Zwicky:1937ec} more than \num{80}~years ago to explain anomalous motions of galaxies gravitationally bound in clusters.  Observational evidence has continued to accumulate since then, including rotation curves of galaxies and their clusters~\cite{Faber:1979em} and discrepancies in the distributions of galaxy cluster mass estimated from luminosity vs. gravitational lensing~\cite{Refregier:2003jl,Clowe:2006hr,Thompson:2015fm}.  That this matter is not only dark but also cold and nonbaryonic is strongly implied by simulations of observed large-scale structure in the Universe~\cite{Springel:2005gv}, fluctuations in the cosmic microwave background radiation~\cite{Komatsu:2011in}, big bang nucleosynthesis~\cite{Copi:1995bx,Cyburt:2016cr}, and analysis of the Lyman-$\alpha$ forest~\cite{Weinberg:2003jn}.

One of the most favored dark matter candidates is the Weakly Interacting Massive Particle (\WIMP)~\cite{Steigman:1985fk,Bertone:2005bi}, which explains the current abundance of dark matter as a thermal relic of the big bang.  Most models predict dark matter WIMP masses near the electroweak scale of \num{100}'s of \si{\GeV\per\square\c}.  However, dark matter particle masses $\leq \SI{10}{\GeV\per\square\c}$ can also be compatible with experimental constraints if a significant asymmetry between dark matter and their anti-particles existed in the early Universe~\cite{Lin:2012bx}.  There are claims of detection or possible detection in this mass range~\cite{Aalseth:2011cg,Agnese:2013hy,Bernabei:2013iw}.

Previous Dark Matter (\DM) searches with \DSf~\cite{Agnes:2015gu,Agnes:2016fz} used pulse shape discrimination (\PSD) on the primary scintillation signals (\SOne) to suppress electron recoil backgrounds, achieving a background-free condition for \DM-induced nuclear recoils (NRs).  Those analyses were sensitive to recoiling argon atoms in the energy range from \DSfUArROIEnergyRange, confining the sensitivity to masses above a few tens of \si{\GeV\per\square\c}.  Here we present a search for \DM\ with a much lower recoil analysis threshold, down to \SI{0.6}{\keVr}, sensitive to \DM\ masses down to \SI{1.8}{\GeV\per\square\c}.  WIMPs in this mass range produce nuclear recoils well below \SI{10}{\keVr}, where the efficiency for detecting the \SOne\ signal is low and \PSD\ is therefore not available.  The required low recoil-energy analysis threshold  is achieved by exploiting the gain inherent in the ionization (\STwo) signal of the dual-phase liquid argon time projection chamber (\LArTPC).  Similar analyses have been presented from dual-phase liquid xenon time projection chambers~\cite{Aprile:2016fv}.

The \DSf\ \LArTPC\ and its veto system are described in Ref.~\cite{Agnes:2015gu}.  The \TPC\ has \DSfPMTsNumber\ \DSfPMTSize\ \PMTs\ (\DSfPMTsNumberTop\ above the transparent anode of the \TPC\ and \DSfPMTsNumberBottom\ below the transparent cathode) viewing a \DSfActiveMass\ active target of low-radioactivity underground argon (\UAr)~\cite{AcostaKane:2008im,Back:2012vo,Back:2012um,Xu:2015do}.  Light signals are detected from both primary \UAr\ scintillation (\SOne) and gas-proportional scintillation (\STwo) from ionization electrons extracted into a vapor layer above the liquid.  The data reported here were acquired between April~30, 2015, and April~25, 2017, using a TPC drift field of \DSfDriftField, an extraction field of \DSfExtractionField, and an electroluminescence (EL)  field of \DSfMultiplicationField.  At this extraction field, the efficiency for extracting ionization electrons into the gas layer is estimated at \DSfElectronExtractionEfficiency~\cite{Bondar:2009gh}. The exposure for the present search including cuts (see below) is \SI{6786.0}{\kg\day}.

The \LArTPC\ lies at the center of a sensitive veto system~\cite{Agnes:2016fw,Agnes:2016cp,Agnes:2017ec}.  The \TPC\ is immersed in a \LSVStainlessSphereDiameter\ diameter liquid scintillator veto (\LSV) filled with \LSVScintillatorMass\  of boron-loaded liquid scintillator and instrumented with \LSVPMTsNumber\ \PMTs.  Surrounding the \LSV\  is a \CTFWaterMass\ Water Cerenkov Veto (\WCV)  instrumented with \CTFPMTsNumber\ \PMTs.  The veto system acts as a highly effective passive shield against local sources of radioactivity. We note that the signals from these detectors are not used in the event analysis because, due to the electron drift time in the TPC, the \STwo\ triggers are not in prompt coincidence with the veto.

The detector has been calibrated {\it in situ} using  $\gamma$ and ($\alpha$,$n$) neutron sources positioned inside the \LSV\ next to the \TPC~\cite{Agnes:2017ec}.  Data taken with \ce{^57Co}, \ce{^133Ba}, and \ce{^137Cs} \gr\ sources are used to validate the Monte Carlo (\MC) simulations, and \AmBe\ and \AmC\ neutron source data are used to verify the nuclear recoil and veto response.  Calibrations were also carried out with \ce{^{83m}Kr} diffused throughout the \TPC\ ~\cite{Lippincott:2010jb}.

We have developed \GFDS~\cite{Agnes:2017cz}, a \Geant-based~\cite{Agostinelli:2003fg,Allison:2006cd} \MC\ code, which describes the performance of the three \DS\ detectors, accounting for material properties, optics, and readout noise. \GFDS~\ also includes a model for \LAr\ scintillation and recombination.  The \MC\ is tuned to agree with the high statistics \ce{^39Ar} data from the  atmospheric argon (\AAr) exposure of \DSf~\cite{Agnes:2015gu}.

Some simple quality cuts were applied to the data before analysis.  Short runs, data where less than the full complement of TPC PMTs was active, and runs with an abnormal trigger rate or with excessive noise on the PMT signal baselines were discarded. 

A hardware event trigger in \DSf\ occurs when  \DSfTriggerPMTsNumberThreshold\ or more PMT  signals exceed a threshold of \DSfTriggerDiscriminatorThreshold\ within a \DSfTriggerWindow\ window. Subsequent triggers are inhibited for \SI{0.8}{ms}, and waveform data are recorded from all 38 PMTs for \SI{440}{\us} starting $\sim$\SI{10}{\us} before the trigger~\cite{Agnes:2017ck}. Software pulse finding algorithms are then applied to the digitized data including the pre-trigger data. The software classifies pulses into two categories (\SOne\ or \STwo) based on the fraction of light detected within the first \WindowFNine\ (\FNine). \SOne\ pulses have \FNine\ values greater than \num{0.15}, as opposed to the much slower-rising \STwo\ pulses.  Unlike previous analyses~\cite{Agnes:2015gu,Agnes:2016fz}, which required both an \SOne\ and an \STwo\ pulse, this analysis achieves a lower energy threshold by accepting not only events with a single \SOne\ and \STwo\ pulse but events with only an \STwo\ pulse.

The efficiency of the software pulse finding algorithm is essentially \DSfPulseFindingEfficiency\ for \STwo\ signals larger than \DSfPulseFindingEfficiencyThreshold~\cite{Agnes:2015gu,Fan:2016ud}.  The pulse finder uses an integration window of \DSfSTwoIntegrationWindow, which is long enough to collect the entire \STwo\ signal including the slow component with its decay time in gas of \DSfSTwoSlowComponentMeanLife~\cite{Agnes:2018vt}.  The pulse integration starts \SI{2}{\us} before the start time of pulses defined by the pulse finding algorithm in order to fully collect the light of slow rise time \STwo\ pulses.

Fiducialization in the present analysis is complicated by the low recoil energy region of interest.  \SOne\ pulses are not usually large enough to be detectable, so no drift time (time between \SOne\ and \STwo\ pulses) is available for $z$-fiducialization.  The usual algorithm for reconstructing the $x$-$y$ position from the \STwo\ light distribution also fails at low recoil energy due to low photoelectron statistics.  Instead, we assign the $x$-$y$ position of each event to be at the center of the \PMT\ receiving the largest number of \STwo\ photoelectrons. We then set a fiducial region in the $x$-$y$ plane by only accepting events where the largest \STwo\ signal is recorded in one of the seven central top-array \PMTs.

We reject a small number of events that have a large \SOne\ pulse, even when accompanied by an abnormally low \STwo\ pulse that would, on its own, fall in the region of interest. These events occur near the wall of the \TPC\ and do not have a regular \STwo\ following them. They are instead accompanied by a small signal from electrons photo-extracted from the cathode by the \SOne\ light. The associated loss of acceptance is much less than \SI{1}{\percent}. Another loss of acceptance due to the misidentification of \STwo\ pulses as \SOne\ pulses is estimated via \GFDS\ simulations to be negligible above the adopted threshold.  This is confirmed by the study of single-electron events discussed below.

\begin{figure}[t!]
\begin{center}
\includegraphics[width=0.5\textwidth]{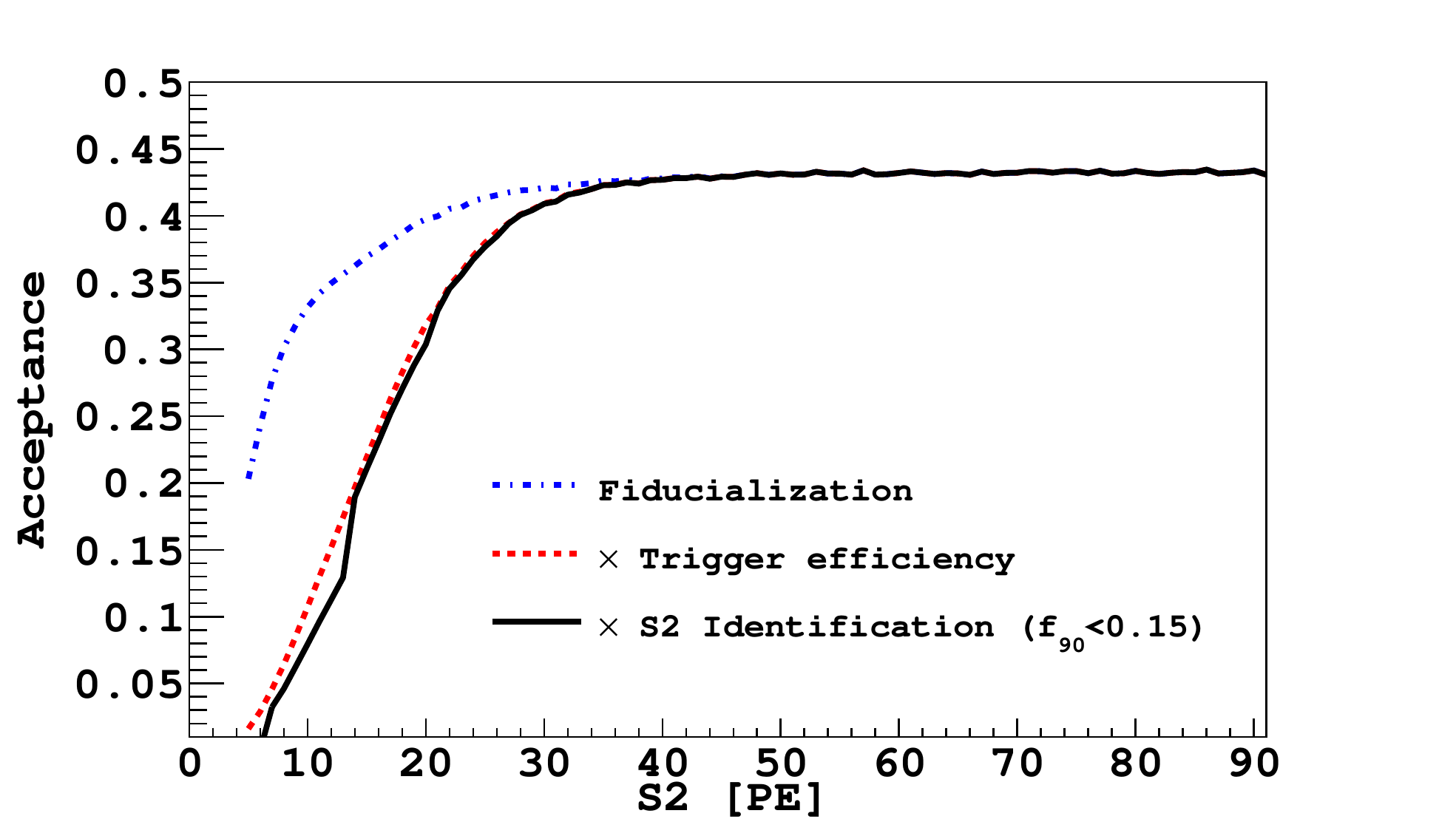}
\end{center}
\caption{\label{fig:DSfEfficiency}Acceptance of the basic cuts described in the text as a function of the number of \si{\pe} in the pulses.}
\end{figure}

The acceptance of the cuts defined above is estimated using a dedicated \MC\ simulation that reproduces the spatial distribution of \STwo\ light predicted by \GFDS~\cite{Agnes:2017cz} and the \STwo\ timing distribution measured in a study of diffusion during electron drift~\cite{Agnes:2018vt}.  \reffig{DSfEfficiency} shows the effect of the above cuts on a sample of  simulated low-energy \STwo-only events that are uniformly distributed throughout the detector.  The figure shows the fraction of events surviving in sequence the fiducial volume cut, the simulated trigger condition, and the \STwo\ identification cut.  The hardware trigger efficiency is \DSfTriggerSTwoEfficiency\ for \STwo\ pulses above \DSfTriggerSTwoEfficiencyThreshold, which is well below the analysis threshold of \Ne\ = \SI{4}{\electron} or recoil energy of \SI{0.1}{\keVee}. The trigger efficiency decreases below this point due to the slow timescale of \STwo\ pulses.  The detector acceptance is \num{0.43(1)} above \SI{30}{\pe} with the dominant acceptance loss due to the restricted fiducial region.  This matches the acceptance of \SI{0.42(1)} found with the same cuts applied to \ce{^39Ar} events from the \DSf\ campaign with an \AAr\ target~\cite{Agnes:2015gu}.

\begin{figure}[t!]
\begin{center}
\includegraphics[width=0.5\textwidth]{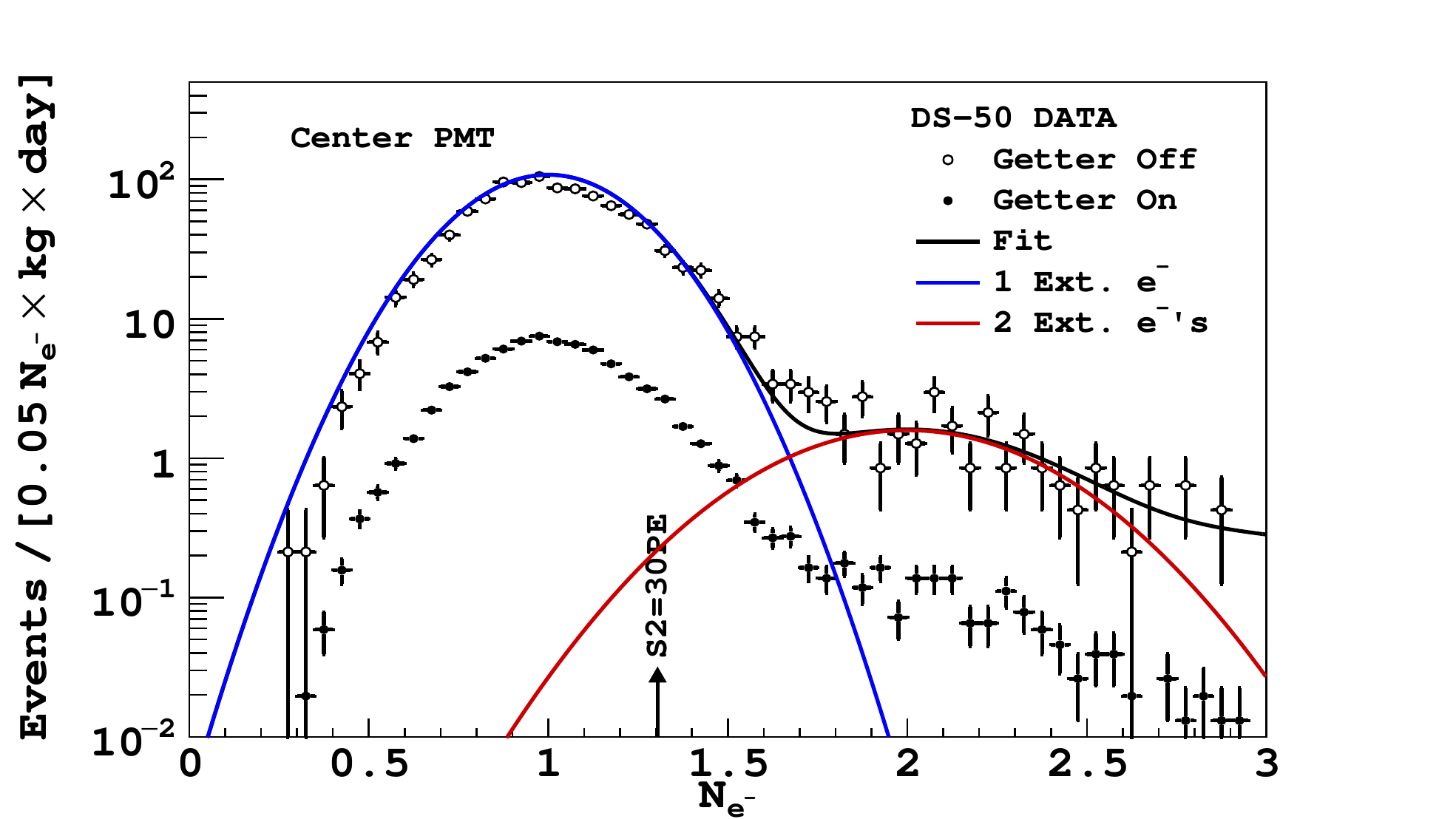}
\end{center}
\caption{\label{fig:NeDistributions} Filled symbols show \DSf\ experimental \Ne\ spectra obtained during the last 100 days of data taking and (open symbols) during the short period where the getter was off for maintenance.  Both the single- and double-electron peaks are seen to be strongly enhanced in the absence of argon purification.  Smooth black curve shows a weighted sum of the \GFDS\ one-, two-, and three-electron responses.}
\end{figure}

The \STwo\ photoelectron yield per extracted ionization electron, $\eta$, is determined by studying single electron events obtained during a short period of time in which the inline argon purification getter was turned off for maintenance purposes (\reffig{NeDistributions}).  These runs have a significantly enhanced single-electron event rate. The observation of strong time and space correlations between single-electron events and preceding large ionization events leads us to believe that these events are from electrons captured by and subsequently released from trace impurities in the argon~\cite{Aprile:2014he,Sorensen:2017wo,Sorensen:2017tf}. We obtain $\eta_{c} = \SI{23(1)}{\pe\per\el}$ for events localized beneath the central PMT, where the error combines variation throughout the entire campaign as well as systematics. 

The rates at which ionization electrons are trapped and subsequently released are found to be \SI{3.5(3)E-5}{\el\per\el} when the getter is off and \SI{0.5(1)E-5}{\el\per\el} when the getter is active normalized to the total yield of ionization electrons.  The electron lifetime was $\sim$\SI{10}{ms} over the entire data taking period, equivalent to $\sim$\SI{30}{ppt} O$_{2}$ contamination.  We ignore data taken where the getter is off and to reduce spurious events from these delayed electrons in standard running, we reject events which occur less than \SI{2.5}{\ms} after a preceding trigger. The resulting loss of exposure is about \SI{1}{\percent}.

Because of an observed radial variation in the electroluminescence yield, a correction is applied to the \STwo\ photoelectron yield for events that originate under the six PMTs surrounding the central one. This correction to the number of extracted electrons, \Ne, was determined using calibrations performed with a mono-energetic (\KrEightThreeQValue) \mbox{$^{83m}$Kr} source to be  $\Ne\ = \STwo / \left(0.76 \cdot \eta_{c} \right)$. 

The \Ne\ distributions expected for different numbers of extracted electrons are modeled with \GFDS\ and are well described by Gaussians. The simulated responses for one and two electrons are in good agreement with the getter-off data.  \reffig{NeDistributions} shows the comparison of the \GFDS\ one- and two-electron distributions with the event distribution in data.

\begin{figure}[t!]
\begin{center}
\includegraphics[width=0.5\textwidth]{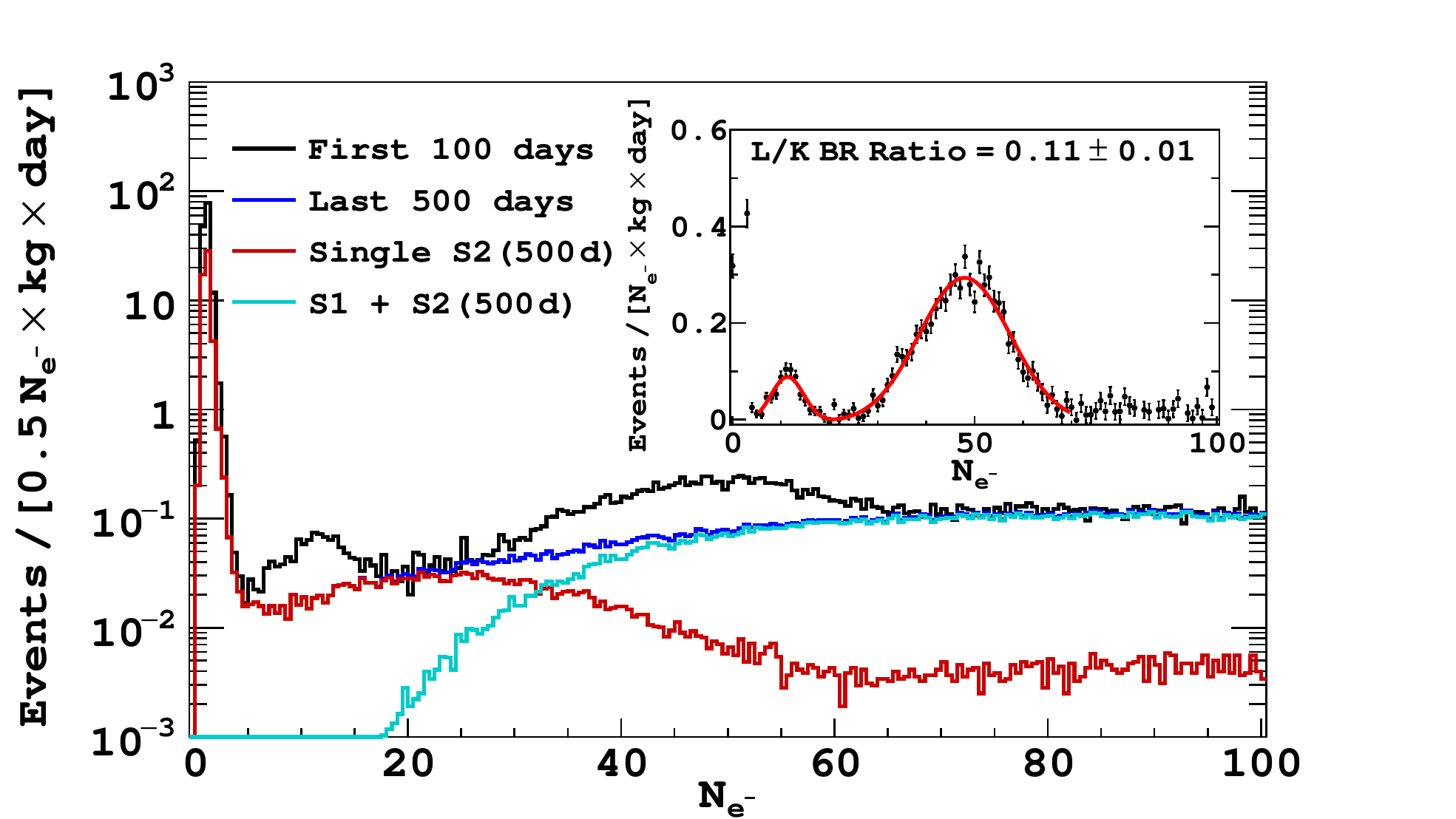}
\end{center}
\caption{\label{fig:37ArSpectrum} Spectrum showing cosmogenic \ce{^37Ar} contributions and their decay as discussed in the text.  Black: first \num{100}~days of present exposure.  Dark blue: last \num{500}~days. Red and cyan show respectively the contributions to the dark blue spectrum from events with only an \STwo\ pulse and from events with a single \SOne\ and a single \STwo\ pulse.  Inset: normalized difference of black minus dark blue, showing the two peaks from \ce{^37Ar} decay.}
\end{figure}

A direct \Ne\ energy calibration for very low energy electron recoils is available from \ce{^37Ar} ($t_{\mbox{\sfrac{1}{2}}} = \ArThreeSevenHalfLife$, EC \SI{100}{\percent}) produced in the \UAr\ by cosmic rays during refining and transport~\cite{Agnes:2016fz}.  \reffig{37ArSpectrum} shows normalized \Ne\ spectra for the first 100~days after the \UAr\ fill and the last 500~days of running, which starts after about 80~days from the end of the 100~days.  The 100-day sample shows two features at \Ne\ around \num{10} and \num{50}, which are shown more clearly in the inset, where the suitably normalized 500-day spectrum has been subtracted.  We attribute these features to the \ArThreeSevenLCaptureXRaysEnergy\ L-shell and the \ArThreeSevenKCaptureXRaysEnergy\ K-shell radiation following electron capture in \ce{^37Ar}~\cite{Kirkwood:1948fg,Pontecorvo:1949kb,Hanna:1949iv}.  These are clearly visible in the first \num{100}~days spectrum and  absent in the remainder of the data set, as expected given the  \ArThreeSevenHalfLife~\cite{Firestone:1999tz} half-life of \ce{^37Ar}.  The observed ratio of the L-shell to the K-shell peak areas is \ArThreeSevenLCapturetoKCaptureRatio, in good agreement with theoretical estimates~\cite{Odiot:1956ha,Brysk:1958ky} and previous experimental results~\cite{SantosOcampo:1960jaa,Sangiorgio:2013gh}.  The widths of the two peaks are consistent with predictions from the \GFDS\ \MC.

The separate contributions from events with a single \STwo\ and those with \SOne+\STwo\ from the 500-day sample are also shown in \reffig{37ArSpectrum}.
The tail of single \STwo\ events extending above \SI{50}{\el}, amounting to about \SI{4}{\percent} of the total rate, is due to unresolved \SOne+\STwo\ events.  These events are mis-categorized but do not affect the total spectral shape. The spike at very low \Ne\ is attributed to electrons trapped by impurities and then released, as discussed above.

{\it In situ} calibration data from \AmC\ and \AmBe\ neutron sources~\cite{Edkins:2017vf} and neutron-beam scattering data from the \SCENE~\cite{Alexander:2013ke,Cao:2015ks} and \ARIS~\cite{Agnes:2018uo} experiments are used to determine the ionization yield from nuclear recoils, \Qy. 

The use of \ce{^241Am} sources for calibration is complicated by the flux of \gr s produced in the sources. In the case of the \AmC\ source, the \gr\ background is reduced by restricting the data to the \num{4} PMTs farthest from the source and the remaining $\gamma$ contamination is estimated using \GFDS, an estimate which is validated by comparison with the data at an energy above any nuclear recoil.  In the case of the \AmBe\ calibration, events in the \TPC\ are accepted only if they were in coincidence with detection of the \SI{4.4}{MeV} $\gamma$ in the veto, a requirement that effectively singles out a pure neutron recoil sample.  The inevitable loss of events (\SI{98}{\percent}) that arises because the signal in the veto is coincident with the \SOne\ signal while the low-energy \STwo\ trigger is delayed by the drift-time in the \TPC\ is manageable given the size of our data-set.

\begin{figure}[t!]
\begin{center}
\includegraphics[width=0.5\textwidth]{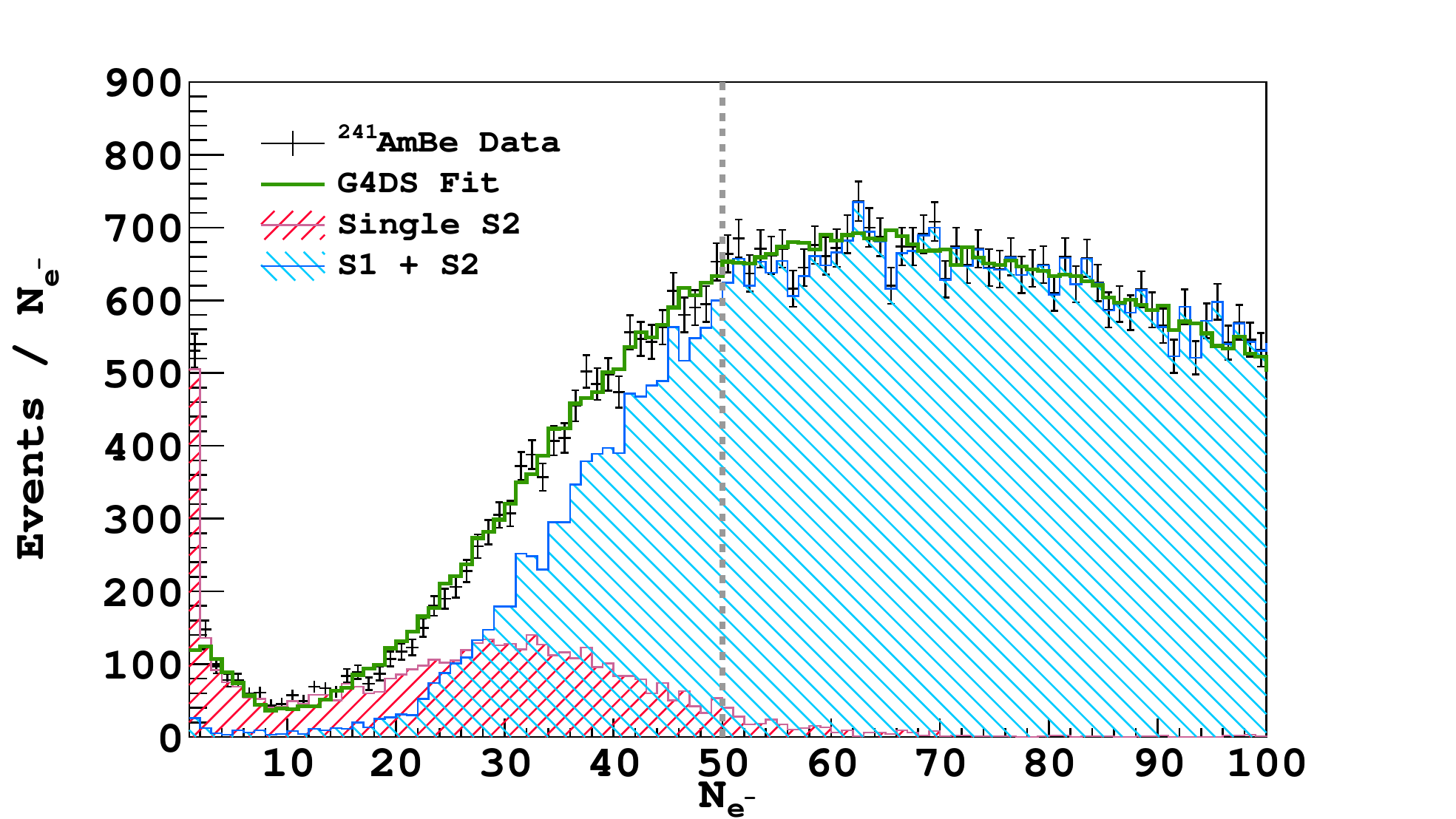}
\end{center}
\caption{\label{fig:AmBeSpectrum} Data and \MC\ fit of the  \Ne\ spectrum for the \AmBe\ run in \DSf. The dashed line shows the lower edge of fit range.}
\end{figure}

\begin{figure}[t!]
\begin{center}
\includegraphics[width=0.5\textwidth]{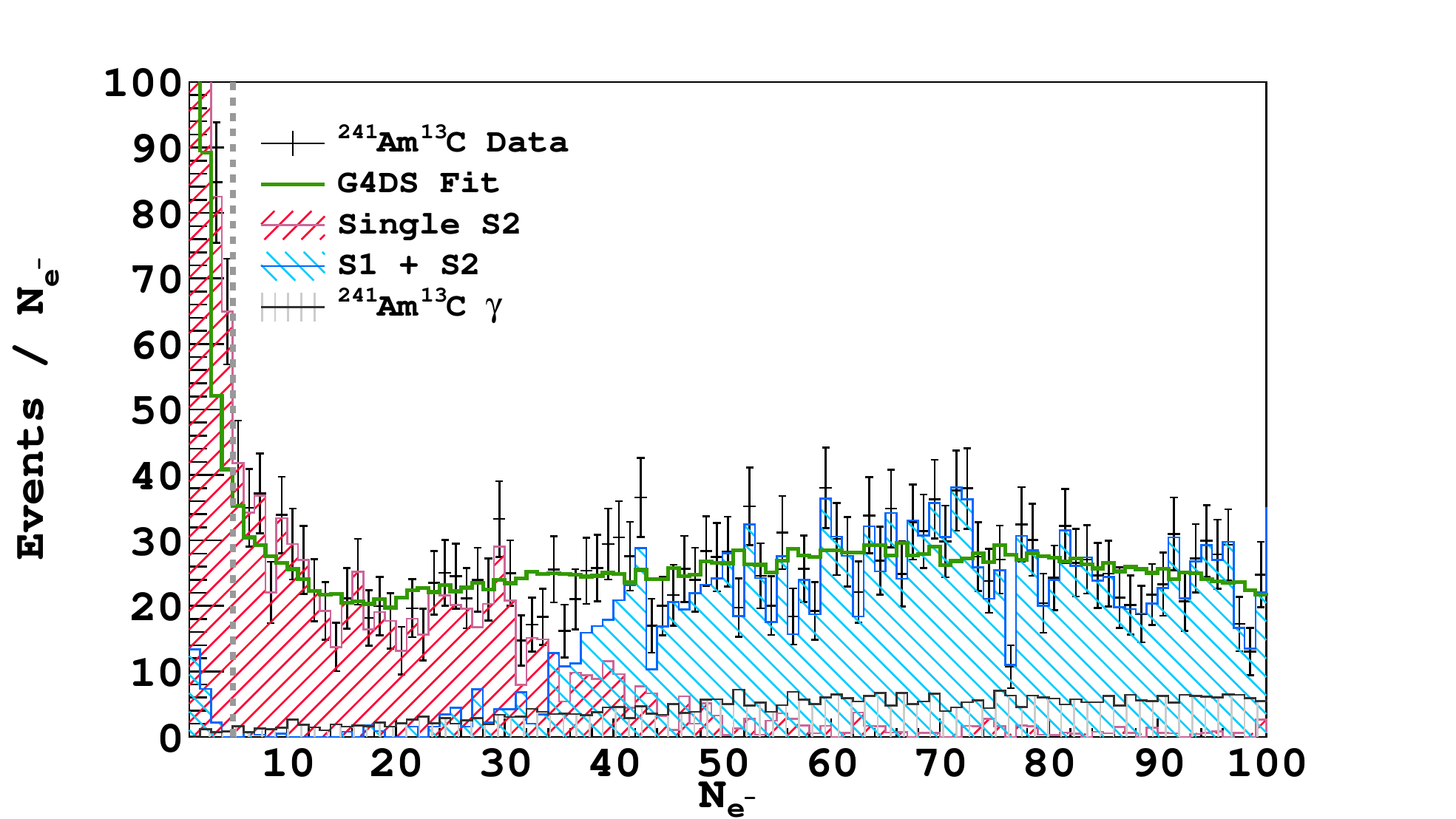}
\end{center}
\caption{\label{fig:AmCSpectrum} Data and \MC\ fit of the \Ne\ spectrum from the \AmC\ run in \DSf.  The dashed line shows the lower edge of fit range.}
\end{figure}

The final \AmC\ and \AmBe\ \Ne\ spectra are fit simultaneously to recoil energy distributions by \GFDS\ using the model of Bezrukov {\it et al.}~\cite{Bezrukov:2011gi} to convert nuclear recoil energy to ionization.  The model has two free parameters that relate to a combination of the energy quenching and the ionization to excitation ratio and the recombination rate of ionization pairs. For the \AmC\ data these two parameters are sufficient and the fit goes to the analysis threshold of \SI{4}{electrons}.  The fit for the \AmBe\ data, however, also includes a term for the acceptance of the coincidence requirement and a strong correlation is noted between the uncertainties on the acceptance-loss model and the ionization response.  To avoid this correlation, the fit to the \AmBe\ data has a threshold of \SI{50}{\electron}, above which the fraction of \STwo\ triggered data is negligible. The resulting fits are shown in \reffig{AmBeSpectrum} and \reffig{AmCSpectrum} for the \AmBe\ and the \AmC\ source, respectively.  The simulated distributions fit the data well and provide strong constraints for the ionization yield.

\begin{figure}[t!]
\begin{center}
\includegraphics[width=0.5\textwidth]{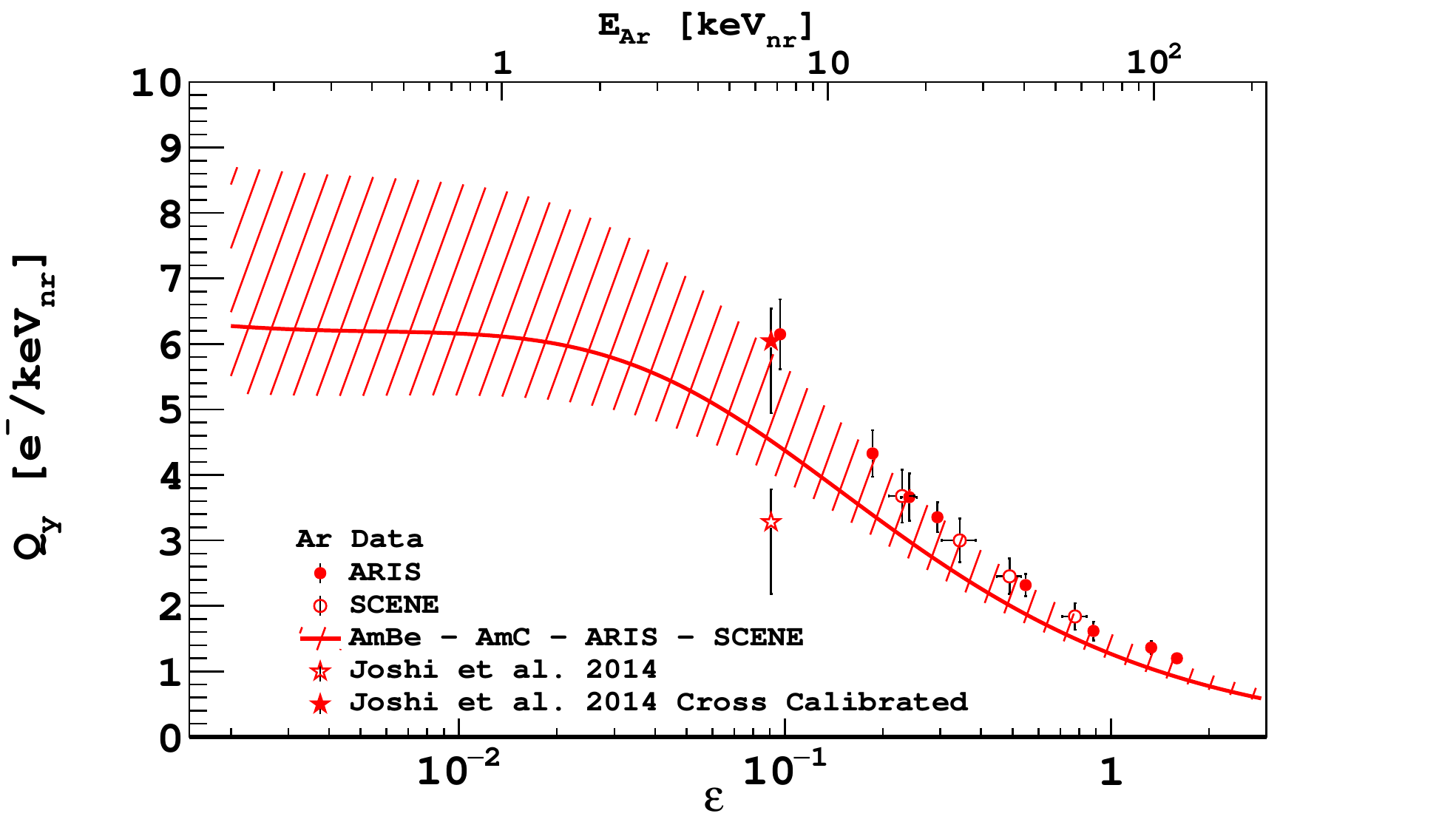}
\end{center}
\caption{\label{fig:NRIonizationYield}The measured ionization yield, \Qy, for nuclear recoils in \LAr\ as a function of the reduced energy parameter, $\epsilon$.  Also shown is the Bezrukov model fit to the \AmBe\ and \AmC\ data (see text).   
}
\end{figure}

\reffig{NRIonizationYield} shows all published ionization yield measurements for argon in our region of interest as a function of $\epsilon$, the reduced energy introduced in Ref.~\cite{Lindhard:1963tr}. Direct measurements of nuclear recoil ionization yield using a neutron-beam were performed by the \SCENE\ experiment~\cite{Alexander:2013ke,Cao:2015ks} and by Joshi {\it et al.}~\cite{Joshi:2014kt,*Joshi:2018pc} at \SI{6.7}{\keVr}.
The measurements of scintillation yield by the \ARIS~\cite{Agnes:2018uo} experiment are converted to ionization yield using the \DSf\ calibration data, where both scintillation and ionization signals are present, and using optical models of both detectors.
The ionization yield from the model fit to the \AmBe\ and \AmC\ data is shown in \reffig{NRIonizationYield} as the solid red curve. The  shaded region below the curve represents the -1 $\sigma$ uncertainty from the fit.  The upper boundary of the shaded region is drawn to represent the ionization predicted using the same model but fitting to the neutron-beam scattering measurements. The difference between the curve and the upper boundary is taken as our systematic uncertainty and is included in the profile likelihood analysis described later.
The ionization yield measured with \AmBe\ and \AmC\ neutron sources in \DSf\ is systematically lower than the ionization yield from \SCENE\ and \ARIS.  The choice of Q$_y$ extracted from \AmBe\ and \AmC\ in this analysis leads to a conservative estimate of the exclusion limits. 


\begin{figure}[t!]
\begin{center}
\includegraphics[width=0.5\textwidth]{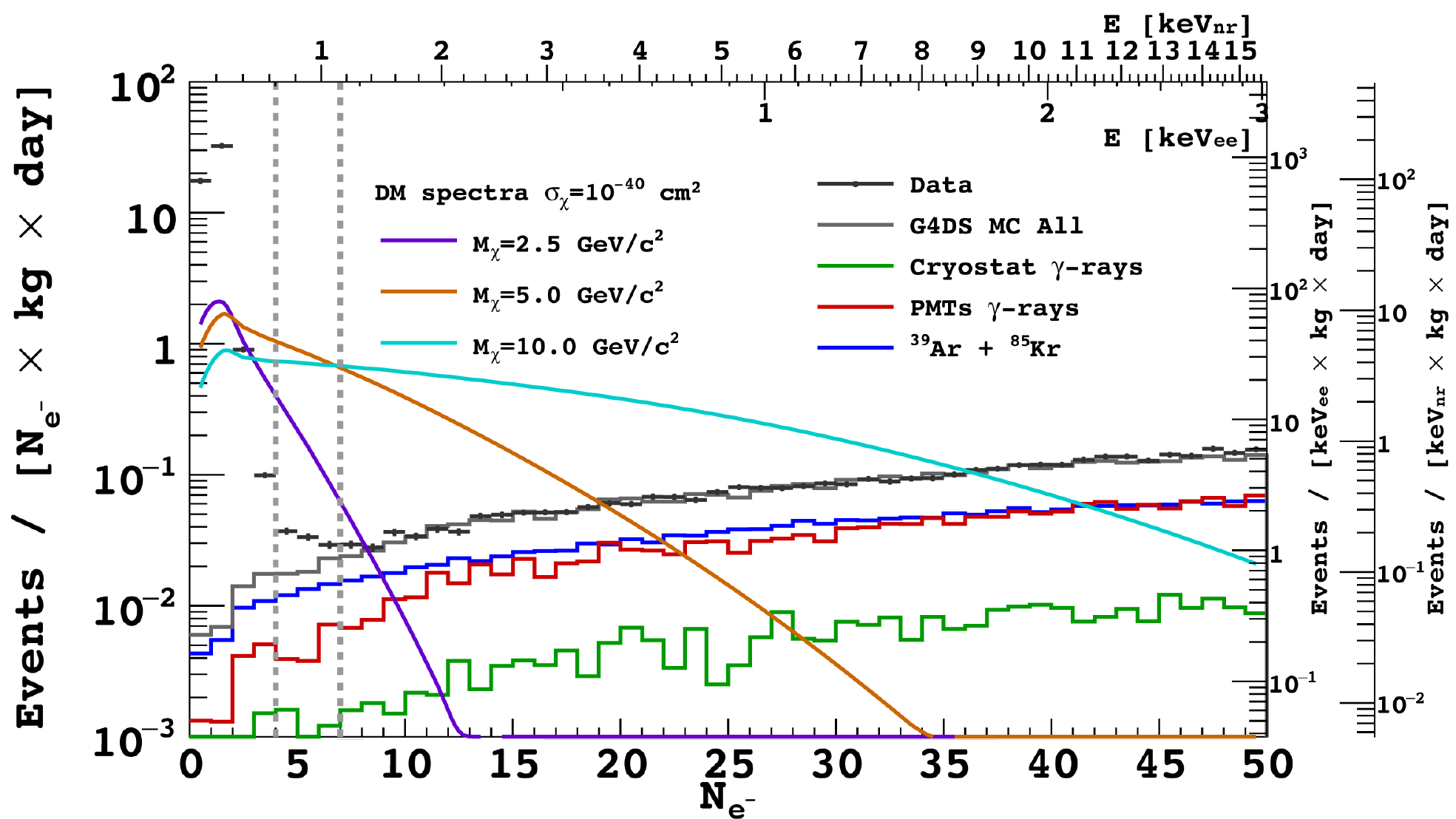}
\end{center}
\caption{\label{fig:NeSpectrum500Days}The \DSf\ \Ne\ spectra at low recoil energy from the analysis of the last \num{500}~days of exposure compared with a \GFDS\ simulation of the background components from known radioactive contaminants.  Also shown are the spectra expected for recoils induced by dark matter particles of masses \num{2.5}, \num{5}, and \SI{10}{\GeV\per\square\c} with a cross section per nucleon of \SI{E-40}{\square\cm} convolved with the no energy quenching fluctuation model and detector resolution.  The $y$-axis scales at right hand side are approximate event rates normalized at $\Ne = \SI{10}{\electron}$.}
\end{figure}

\reffig{NeSpectrum500Days} shows the \Ne\ spectrum for the last \num{500}~days (same as blue histogram in \reffig{37ArSpectrum}) together with the contributions from the individual radiation sources from the simulation, normalized using  the detector construction materials radioassay data and radioactivity estimation obtained by fitting gamma lines at high energy, \ce{^39Ar}, and \ce{^85Kr} spectra.  The \Ne\ distribution from the \num{500}~day sample obtained with the present analysis is consistent within uncertainties with the \GFDS\ \MC\ simulation~\cite{Agnes:2016fz,Agnes:2017cz} for $\Ne \gtrsim \SI{7}{\electron}$ ($\sim\SI{1}{\keVr}$).  
There is an excess of data in the region of \Ne\ of \SIrange{4}{7}{\electron}, the origin of which is left for future study.

The  observed \DSf\ rate as a function of \SI{}{\keVee} is flat at \SI{\sim1.5}{\ev\per\keVee\per\kg\per\day} in the range from \SIrange{0.1}{10}{\keVee}.  The large (\num{E2}) increase below \SI{0.1}{\keVee} is believed to be from electrons trapped and subsequently released by impurities.  This is based on the observation of a strong time correlation between a higher energy event and the following low \Ne\ events, suggesting electrons are released from impurities with a $\sim$\SI{50}{ms} time constant.  Also shown in \reffig{NeSpectrum500Days} are the \Ne\ spectra expected for nuclear recoils induced by dark matter particles of masses \num{2.5}, \num{5}, and \SI{10}{\GeV\per\square\c} with a cross section of \SI{E-40}{\square\cm} and standard isothermal halo parameters ($v_{\text{escape}} = \SI{544}{\km\per\sec}$, $v_{0} = \SI{220}{\km\per\sec}$, $v_{\text{Earth}} = \SI{232}{\km\per\sec}$, and $\rho_{\text{DM}} = \SI{0.3}{\GeV\per\square\c\per\cubic\cm}$~\cite{Smith:2007fi}).

Uncertainties in the expected signal yield above the analysis threshold are dominated by the average ionization yield as extracted from the \AmBe\ and \AmC\ data and its intrinsic fluctuations.  We have no {\it a priori} knowledge of the width of the ionization distribution of nuclear recoils and are not aware of measurements in liquid argon in the energy range of interest.  We therefore consider two extreme models: one allowing for fluctuations in energy quenching, ionization yield, and recombination processes obtained with binomial distributions and another where the fluctuations in energy quenching are set to zero, equivalent to imposing an analysis threshold of \SI{0.59}{\keVr}.

Extrapolations of the expected background to the signal region are mostly affected by theoretical uncertainties on the low energy portion of the \ce{^85Kr} and \ce{^39Ar} $\beta$-spectra and by the uncertainty in the electron recoil energy scale and resolution.

Upper limits on the WIMP-nucleon scattering cross-section are extracted from the observed \Ne\ spectrum using a binned profile likelihood method~\cite{Cowan:2011cx,Moneta:2011jc,Verkerke:2006es}.  Two signal regions are defined, the first one using a threshold of \SI{4}{\electron}, determined by the approximate end of the trapped electron background spectrum, and the second above a threshold of \SI{7}{\electron}, where the background is described within uncertainties by the \GFDS\ simulation.  The first region has sensitivity to the entire range of \DM\ masses explored in this work, but the data is contaminated by a component  that is not included in the background model, resulting in weaker bounds on the \DM-nucleon cross-section.  The second signal region has limited sensitivity to \DM\ masses below \SI{3.5}{\GeV\per\square\c} but, due to the agreement between data and background model, more tightly constrains the cross-section at higher masses.  For a given fluctuation model and \DM\ mass, we calculate limits using both signal regions and quote the more stringent of the two.

\begin{figure}[t!]
\begin{center}
\includegraphics[width=0.5\textwidth]{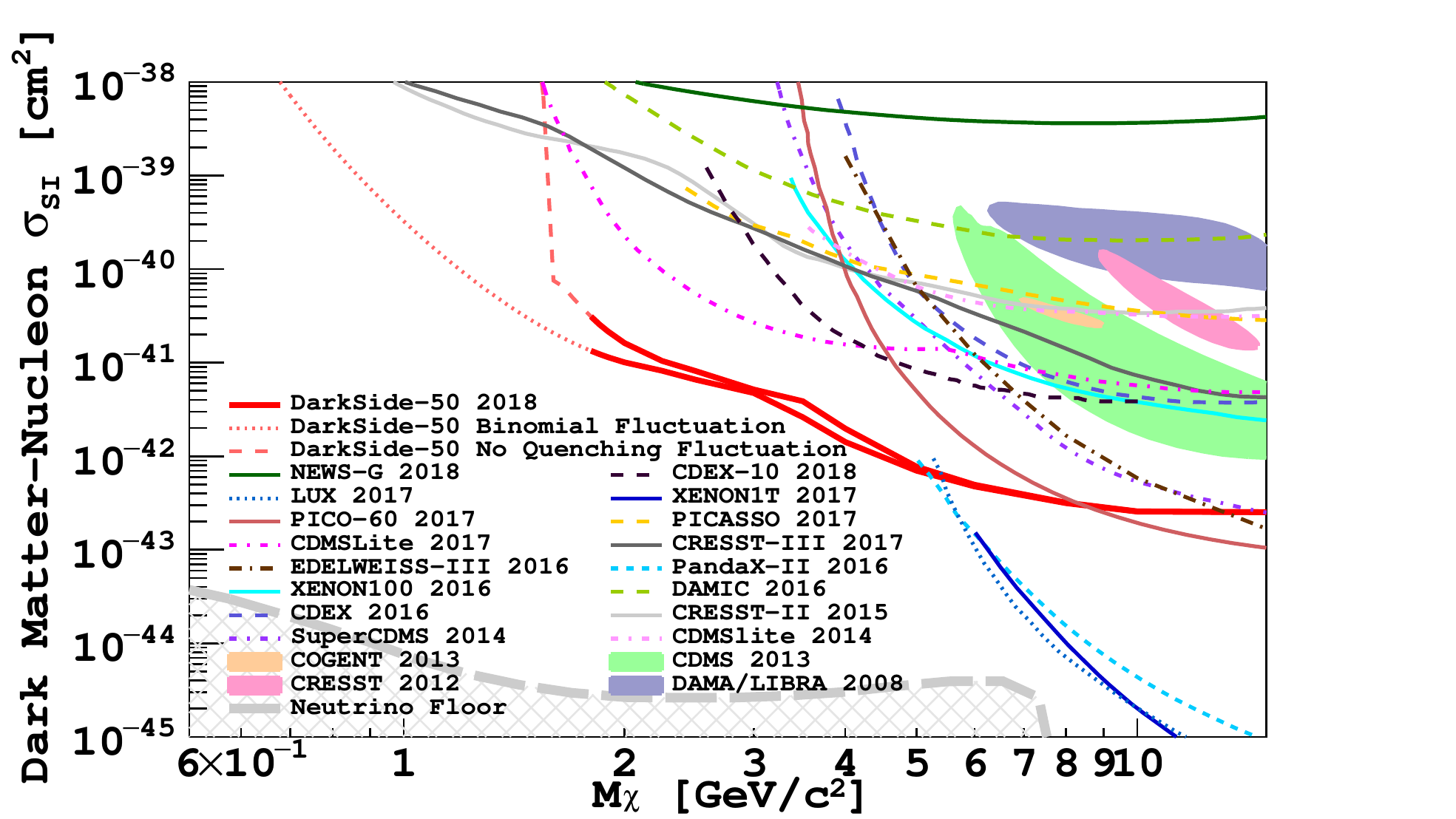}
\end{center}
\caption{\label{fig:ExclusionLimits}\SI{90}{\percent} upper limits on spin independent \DM-nucleon cross sections from \DSf\ in the range above \SI{1.8}{\GeV\per\square\c}.  See the text for additional details.}
\end{figure}

The \SI{90}{\percent} C.L. exclusion curves for the binomial fluctuation model (red dotted line) and the model with zero fluctuation in the energy quenching (red dashed line) are shown in \reffig{ExclusionLimits}.  For masses above \SI{1.8}{\GeV\per\square\c}, the \SI{90}{\percent} C.L. exclusion is nearly insensitive to the choice of quenching fluctuation model.   Below \SI{1.8}{\GeV\per\square\c}, the two exclusion curves rapidly diverge because of the effective threshold due to the absence of the fluctuations in the energy quenching process.
Without additional constraints on the quenching fluctuations, it is impossible to claim an exclusion in this mass range.

Our exclusion limit above \SI{1.8}{\GeV\per\square\c} is compared with the \SI{90}{\percent} C.L. exclusion limits from Refs.~\cite{Agnese:2014cq,Agnese:2014ix,Angloher:2016js,Zhao:2016bq,Hehn:2016eq,AguilarArevalo:2016ei,Aprile:2016fv,Tan:2016fv,Petricca:2017vx,Agnese:2017wy,Behnke:2017cc,Amole:2017iz,Aprile:2017eq,Akerib:2017kg,Arnaud:2018fg,Jiang:2018ez}, the region of claimed discovery of Refs.~\cite{Savage:2009hi,Angloher:2012kl,Agnese:2013hy,Aalseth:2013bg}, and the neutrino floor for \LAr\ experiments~\cite{Ruppin:2014ee}.  Improved ionization yield measurement and assessment of a realistic ionization fluctuation model, which are left for future work, may be used to determine the actual sensitivity of the present experiment within the range indicated by the two curves below the \SI{1.8}{\GeV\per\square\c} \DM\ mass.

\begin{acknowledgments}
The \DS\ Collaboration offers its profound gratitude to the \LNGS\ and its staff for their invaluable technical and logistical support.  We also thank the Fermilab Particle Physics, Scientific, and Core Computing Divisions.  Construction and operation of the \DSf\ detector was supported by the U.S. National Science Foundation (NSF) (Grants \grant{PHY}{0919363}, \grant{PHY}{1004072}, \grant{PHY}{1004054}, \grant{PHY}{1242585}, \grant{PHY}{1314483}, \grant{PHY}{1314501}, \grant{PHY}{1314507}, \grant{PHY}{1352795}, \grant{PHY}{1622415}, and associated collaborative grants \grant{PHY}{1211308} and \grant{PHY}{1455351}), the Italian Istituto Nazionale di Fisica Nucleare, the U.S. Department of Energy (Contracts \grant{DE}{FG02-91ER40671}, \grant{DE}{AC02-07CH11359}, and \grant{DE}{AC05-76RL01830}), the Russian Science Foundation (Grant \grant{16}{12-10369}), the Polish NCN (Grant \grant{UMO}{2014/15/B/ST2/02561}) and the Foundation for Polish Science (Grant \grant{Team2016}{2/17}).  We also acknowledge financial support from the FrenchÊInstitut National de Physique Nucl\'eaire et de Physique des Particules (IN2P3), from the UnivEarthS Labex program of Sorbonne Paris Cit\'e (Grants \grant{ANR}{10-LABX-0023} and \grant{ANR}{11-IDEX-0005-02}), and from the S\~ao Paulo Research Foundation (FAPESP) (Grant \grant{2016/09084}{0}).
\end{acknowledgments}

\bibliographystyle{$TEXMFHOME/ds}
\bibliography{$TEXMFHOME/ds,privatecommunicationJoshi}

\begin{thebibliography}{78}
\expandafter\ifx\csname natexlab\endcsname\relax\def\natexlab#1{#1}\fi
\expandafter\ifx\csname bibnamefont\endcsname\relax
  \def\bibnamefont#1{#1}\fi
\expandafter\ifx\csname bibfnamefont\endcsname\relax
  \def\bibfnamefont#1{#1}\fi
\expandafter\ifx\csname citenamefont\endcsname\relax
  \def\citenamefont#1{#1}\fi
\expandafter\ifx\csname url\endcsname\relax
  \def\url#1{\texttt{#1}}\fi
\expandafter\ifx\csname urlprefix\endcsname\relax\def\urlprefix{URL }\fi
\providecommand{\bibinfo}[2]{#2}
\providecommand{\eprint}[2][]{\url{#2}}

\bibitem[{\citenamefont{Oort}(1932)}]{Oort:1932tb}
\bibinfo{author}{\bibfnamefont{J.~H.} \bibnamefont{Oort}},
  \href{http://adsabs.harvard.edu/abs/1932BAN.....6..249O}{\bibinfo{journal}{Bull.
  Astron. Inst. Netherlands} \textbf{\bibinfo{volume}{6}},
  \bibinfo{pages}{249}\bibinfo{year}{ (\bibinfo{year}{1932})}}.

\bibitem[{\citenamefont{Zwicky}(1933)}]{Zwicky:1933ub}
\bibinfo{author}{\bibfnamefont{F.}~\bibnamefont{Zwicky}},
  \href{http://adsabs.harvard.edu/abs/1933AcHPh...6..110Z}{\bibinfo{journal}{Helv.
  Phys. Acta} \textbf{\bibinfo{volume}{6}}, \bibinfo{pages}{110}\bibinfo{year}{
  (\bibinfo{year}{1933})}}.

\bibitem[{\citenamefont{Zwicky}(1937)}]{Zwicky:1937ec}
\bibinfo{author}{\bibfnamefont{F.}~\bibnamefont{Zwicky}},
  \href{http://dx.doi.org/10.1086/143864}{\bibinfo{journal}{Ap. J.}
  \textbf{\bibinfo{volume}{86}}, \bibinfo{pages}{217}\bibinfo{year}{
  (\bibinfo{year}{1937})}}.

\bibitem[{\citenamefont{Faber and Gallagher}(1979)}]{Faber:1979em}
\bibinfo{author}{\bibfnamefont{S.~M.} \bibnamefont{Faber}} \bibnamefont{and}
  \bibinfo{author}{\bibfnamefont{J.~S.} \bibnamefont{Gallagher}},
  \href{http://dx.doi.org/10.1146/annurev.aa.17.090179.001031}{\bibinfo{journal}{Annu.
  Rev. Astro. Astrophys.} \textbf{\bibinfo{volume}{17}},
  \bibinfo{pages}{135}\bibinfo{year}{ (\bibinfo{year}{1979})}}.

\bibitem[{\citenamefont{Refregier}(2003)}]{Refregier:2003jl}
\bibinfo{author}{\bibfnamefont{A.}~\bibnamefont{Refregier}},
  \href{http://dx.doi.org/10.1146/annurev.astro.41.111302.102207}{\bibinfo{journal}{Annu.
  Rev. Astro. Astrophys.} \textbf{\bibinfo{volume}{41}},
  \bibinfo{pages}{645}\bibinfo{year}{ (\bibinfo{year}{2003})}}.

\bibitem[{\citenamefont{Clowe et~al.}(2006)}]{Clowe:2006hr}
\bibinfo{author}{\bibfnamefont{D.}~\bibnamefont{Clowe}}  \bibnamefont{et~al.},
  \href{http://dx.doi.org/10.1086/508162}{\bibinfo{journal}{Ap. J.}
  \textbf{\bibinfo{volume}{648}}, \bibinfo{pages}{L109}\bibinfo{year}{
  (\bibinfo{year}{2006})}}.

\bibitem[{\citenamefont{Thompson et~al.}(2015)\citenamefont{Thompson, Dav{\'e},
  and Nagamine}}]{Thompson:2015fm}
\bibinfo{author}{\bibfnamefont{R.}~\bibnamefont{Thompson}},
  \bibinfo{author}{\bibfnamefont{R.}~\bibnamefont{Dav{\'e}}},
  \bibnamefont{and} \bibinfo{author}{\bibfnamefont{K.}~\bibnamefont{Nagamine}},
   \href{http://dx.doi.org/10.1093/mnras/stv1433}{\bibinfo{journal}{Month. Not.
  Royal Astron. Soc.} \textbf{\bibinfo{volume}{452}},
  \bibinfo{pages}{3030}\bibinfo{year}{ (\bibinfo{year}{2015})}}.

\bibitem[{\citenamefont{Springel et~al.}(2005)}]{Springel:2005gv}
\bibinfo{author}{\bibfnamefont{V.}~\bibnamefont{Springel}}
  \bibnamefont{et~al.},
  \href{http://dx.doi.org/10.1038/nature03597}{\bibinfo{journal}{Nature}
  \textbf{\bibinfo{volume}{435}}, \bibinfo{pages}{629}\bibinfo{year}{
  (\bibinfo{year}{2005})}}.

\bibitem[{\citenamefont{Komatsu et~al.}(2011)}]{Komatsu:2011in}
\bibinfo{author}{\bibfnamefont{E.}~\bibnamefont{Komatsu}}  \bibnamefont{et~al.}
  (\bibinfo{affiliation}{The WMAP Collaboration}),
  \href{http://dx.doi.org/10.1088/0067-0049/192/2/18}{\bibinfo{journal}{Ap. J.
  Supp. Ser.} \textbf{\bibinfo{volume}{192}},
  \bibinfo{pages}{18}\bibinfo{year}{ (\bibinfo{year}{2011})}}.

\bibitem[{\citenamefont{Copi et~al.}(1995)\citenamefont{Copi, Schramm, and
  Turner}}]{Copi:1995bx}
\bibinfo{author}{\bibfnamefont{C.}~\bibnamefont{Copi}},
  \bibinfo{author}{\bibfnamefont{D.}~\bibnamefont{Schramm}},  \bibnamefont{and}
  \bibinfo{author}{\bibfnamefont{M.}~\bibnamefont{Turner}}
  (\bibinfo{affiliation}{Department of Physics, University of Chicago, IL
  60637-1433.}),
  \href{http://dx.doi.org/10.1126/science.7809624}{\bibinfo{journal}{Science}
  \textbf{\bibinfo{volume}{267}}, \bibinfo{pages}{192}\bibinfo{year}{
  (\bibinfo{year}{1995})}}.

\bibitem[{\citenamefont{Cyburt et~al.}(2016)\citenamefont{Cyburt, Fields,
  Olive, and Yeh}}]{Cyburt:2016cr}
\bibinfo{author}{\bibfnamefont{R.~H.} \bibnamefont{Cyburt}},
  \bibinfo{author}{\bibfnamefont{B.~D.} \bibnamefont{Fields}},
  \bibinfo{author}{\bibfnamefont{K.~A.} \bibnamefont{Olive}},
  \bibnamefont{and} \bibinfo{author}{\bibfnamefont{T.-H.} \bibnamefont{Yeh}},
  \href{http://dx.doi.org/10.1103/RevModPhys.88.015004}{\bibinfo{journal}{Rev.
  Mod. Phys.} \textbf{\bibinfo{volume}{88}},
  \bibinfo{pages}{461}\bibinfo{year}{ (\bibinfo{year}{2016})}}.

\bibitem[{\citenamefont{Weinberg}(2003)}]{Weinberg:2003jn}
\bibinfo{author}{\bibfnamefont{D.~H.} \bibnamefont{Weinberg}},
  \href{http://dx.doi.org/10.1063/1.1581786}{\bibinfo{journal}{AIP Conf. Proc.}
  \textbf{\bibinfo{volume}{666}}, \bibinfo{pages}{157}\bibinfo{year}{
  (\bibinfo{year}{2003})}}.

\bibitem[{\citenamefont{Steigman and Turner}(1985)}]{Steigman:1985fk}
\bibinfo{author}{\bibfnamefont{G.}~\bibnamefont{Steigman}} \bibnamefont{and}
  \bibinfo{author}{\bibfnamefont{M.~S.} \bibnamefont{Turner}},
  \href{http://dx.doi.org/10.1016/0550-3213(85)90537-1}{\bibinfo{journal}{Nucl.
  Phys. B} \textbf{\bibinfo{volume}{253}}, \bibinfo{pages}{375}\bibinfo{year}{
  (\bibinfo{year}{1985})}}.

\bibitem[{\citenamefont{Bertone et~al.}(2005)\citenamefont{Bertone, Hooper, and
  Silk}}]{Bertone:2005bi}
\bibinfo{author}{\bibfnamefont{G.}~\bibnamefont{Bertone}},
  \bibinfo{author}{\bibfnamefont{D.}~\bibnamefont{Hooper}},  \bibnamefont{and}
  \bibinfo{author}{\bibfnamefont{J.}~\bibnamefont{Silk}},
  \href{http://dx.doi.org/10.1016/j.physrep.2004.08.031}{\bibinfo{journal}{Phys.
  Rep.} \textbf{\bibinfo{volume}{405}}, \bibinfo{pages}{279}\bibinfo{year}{
  (\bibinfo{year}{2005})}}.

\bibitem[{\citenamefont{Lin et~al.}(2012)\citenamefont{Lin, Yu, and
  Zurek}}]{Lin:2012bx}
\bibinfo{author}{\bibfnamefont{T.}~\bibnamefont{Lin}},
  \bibinfo{author}{\bibfnamefont{H.-B.} \bibnamefont{Yu}},  \bibnamefont{and}
  \bibinfo{author}{\bibfnamefont{K.~M.} \bibnamefont{Zurek}},
  \href{http://dx.doi.org/10.1103/PhysRevD.85.063503}{\bibinfo{journal}{Phys.
  Rev. D} \textbf{\bibinfo{volume}{85}}, \bibinfo{pages}{063503}\bibinfo{year}{
  (\bibinfo{year}{2012})}}.

\bibitem[{\citenamefont{Aalseth et~al.}(2011)}]{Aalseth:2011cg}
\bibinfo{author}{\bibfnamefont{C.~E.} \bibnamefont{Aalseth}}
  \bibnamefont{et~al.} (\bibinfo{affiliation}{The CoGeNT Collaboration}),
  \href{http://dx.doi.org/10.1103/PhysRevLett.106.131301}{\bibinfo{journal}{Phys.
  Rev. Lett.} \textbf{\bibinfo{volume}{106}},
  \bibinfo{pages}{131301}\bibinfo{year}{ (\bibinfo{year}{2011})}}.

\bibitem[{\citenamefont{Agnese et~al.}(2013)}]{Agnese:2013hy}
\bibinfo{author}{\bibfnamefont{R.}~\bibnamefont{Agnese}}  \bibnamefont{et~al.}
  (\bibinfo{affiliation}{The CDMS Collaboration}),
  \href{http://dx.doi.org/10.1103/PhysRevLett.111.251301}{\bibinfo{journal}{Phys.
  Rev. Lett.} \textbf{\bibinfo{volume}{111}},
  \bibinfo{pages}{251301}\bibinfo{year}{ (\bibinfo{year}{2013})}}.

\bibitem[{\citenamefont{Bernabei et~al.}(2013)}]{Bernabei:2013iw}
\bibinfo{author}{\bibfnamefont{R.}~\bibnamefont{Bernabei}}
  \bibnamefont{et~al.},
  \href{http://dx.doi.org/10.1140/epjc/s10052-013-2648-7}{\bibinfo{journal}{Eur.
  Phys. J. C} \textbf{\bibinfo{volume}{73}},
  \bibinfo{pages}{297}\bibinfo{year}{ (\bibinfo{year}{2013})}}.

\bibitem[{\citenamefont{Agnes et~al.}(2015)}]{Agnes:2015gu}
\bibinfo{author}{\bibfnamefont{P.}~\bibnamefont{Agnes}}  \bibnamefont{et~al.}
  (\bibinfo{affiliation}{The DarkSide Collaboration}),
  \href{http://dx.doi.org/10.1016/j.physletb.2015.03.012}{\bibinfo{journal}{Phys.
  Lett. B} \textbf{\bibinfo{volume}{743}}, \bibinfo{pages}{456}\bibinfo{year}{
  (\bibinfo{year}{2015})}}.

\bibitem[{\citenamefont{Agnes et~al.}(2016{\natexlab{a}})}]{Agnes:2016fz}
\bibinfo{author}{\bibfnamefont{P.}~\bibnamefont{Agnes}}  \bibnamefont{et~al.}
  (\bibinfo{affiliation}{The DarkSide Collaboration}),
  \href{http://dx.doi.org/10.1103/PhysRevD.93.081101}{\bibinfo{journal}{Phys.
  Rev. D} \textbf{\bibinfo{volume}{93}}, \bibinfo{pages}{081101}\bibinfo{year}{
  (\bibinfo{year}{2016}{\natexlab{a}})}}.

\bibitem[{\citenamefont{Aprile et~al.}(2016)}]{Aprile:2016fv}
\bibinfo{author}{\bibfnamefont{E.}~\bibnamefont{Aprile}}  \bibnamefont{et~al.}
  (\bibinfo{affiliation}{The XENON Collaboration}),
  \href{http://dx.doi.org/10.1103/PhysRevD.94.092001}{\bibinfo{journal}{Phys.
  Rev. D} \textbf{\bibinfo{volume}{94}}, \bibinfo{pages}{092001}\bibinfo{year}{
  (\bibinfo{year}{2016})}}.

\bibitem[{\citenamefont{Acosta-Kane et~al.}(2008)}]{AcostaKane:2008im}
\bibinfo{author}{\bibfnamefont{D.}~\bibnamefont{Acosta-Kane}}
  \bibnamefont{et~al.},
  \href{http://dx.doi.org/10.1016/j.nima.2007.12.032}{\bibinfo{journal}{Nucl.
  Inst. Meth. A} \textbf{\bibinfo{volume}{587}},
  \bibinfo{pages}{46}\bibinfo{year}{ (\bibinfo{year}{2008})}}.

\bibitem[{\citenamefont{Back et~al.}(2012{\natexlab{a}})}]{Back:2012vo}
\bibinfo{author}{\bibfnamefont{H.~O.} \bibnamefont{Back}}
  \bibnamefont{et~al.},
  \href{http://arxiv.org/abs/1204.6024v2}{\bibinfo{journal}{arXiv}:1204.6024v2\bibinfo{year}{
  (\bibinfo{year}{2012}{\natexlab{a}})}}.

\bibitem[{\citenamefont{Back et~al.}(2012{\natexlab{b}})}]{Back:2012um}
\bibinfo{author}{\bibfnamefont{H.~O.} \bibnamefont{Back}}
  \bibnamefont{et~al.},
  \href{http://arxiv.org/abs/1204.6061v2}{\bibinfo{journal}{arXiv}:1204.6061v2\bibinfo{year}{
  (\bibinfo{year}{2012}{\natexlab{b}})}}.

\bibitem[{\citenamefont{Xu et~al.}(2015)}]{Xu:2015do}
\bibinfo{author}{\bibfnamefont{J.}~\bibnamefont{Xu}}  \bibnamefont{et~al.},
  \href{http://dx.doi.org/10.1016/j.astropartphys.2015.01.002}{\bibinfo{journal}{Astropart.
  Phys.} \textbf{\bibinfo{volume}{66}}, \bibinfo{pages}{53}\bibinfo{year}{
  (\bibinfo{year}{2015})}}.

\bibitem[{\citenamefont{Bondar et~al.}(2009)}]{Bondar:2009gh}
\bibinfo{author}{\bibfnamefont{A.}~\bibnamefont{Bondar}}  \bibnamefont{et~al.},
   \href{http://dx.doi.org/10.1088/1748-0221/4/09/P09013}{\bibinfo{journal}{JINST}
  \textbf{\bibinfo{volume}{4}}, \bibinfo{pages}{P09013}\bibinfo{year}{
  (\bibinfo{year}{2009})}}.

\bibitem[{\citenamefont{Agnes et~al.}(2016{\natexlab{b}})}]{Agnes:2016fw}
\bibinfo{author}{\bibfnamefont{P.}~\bibnamefont{Agnes}}  \bibnamefont{et~al.}
  (\bibinfo{affiliation}{The DarkSide collaboration}),
  \href{http://dx.doi.org/10.1088/1748-0221/11/03/P03016}{\bibinfo{journal}{JINST}
  \textbf{\bibinfo{volume}{11}}, \bibinfo{pages}{P03016}\bibinfo{year}{
  (\bibinfo{year}{2016}{\natexlab{b}})}}.

\bibitem[{\citenamefont{Agnes et~al.}(2016{\natexlab{c}})}]{Agnes:2016cp}
\bibinfo{author}{\bibfnamefont{P.}~\bibnamefont{Agnes}}  \bibnamefont{et~al.}
  (\bibinfo{affiliation}{The DarkSide Collaboration}),
  \href{http://dx.doi.org/10.1088/1748-0221/11/12/P12007}{\bibinfo{journal}{JINST}
  \textbf{\bibinfo{volume}{11}}, \bibinfo{pages}{P12007}\bibinfo{year}{
  (\bibinfo{year}{2016}{\natexlab{c}})}}.

\bibitem[{\citenamefont{Agnes et~al.}(2017{\natexlab{a}})}]{Agnes:2017ec}
\bibinfo{author}{\bibfnamefont{P.}~\bibnamefont{Agnes}}  \bibnamefont{et~al.}
  (\bibinfo{affiliation}{The DarkSide Collaboration}),
  \href{http://dx.doi.org/10.1088/1748-0221/12/12/T12004}{\bibinfo{journal}{JINST}
  \textbf{\bibinfo{volume}{12}}, \bibinfo{pages}{T12004}\bibinfo{year}{
  (\bibinfo{year}{2017}{\natexlab{a}})}}.

\bibitem[{\citenamefont{Lippincott et~al.}(2010)}]{Lippincott:2010jb}
\bibinfo{author}{\bibfnamefont{W.~H.} \bibnamefont{Lippincott}}
  \bibnamefont{et~al.},
  \href{http://dx.doi.org/10.1103/PhysRevC.81.045803}{\bibinfo{journal}{Phys.
  Rev. C} \textbf{\bibinfo{volume}{81}}, \bibinfo{pages}{045803}\bibinfo{year}{
  (\bibinfo{year}{2010})}}.

\bibitem[{\citenamefont{Agnes et~al.}(2017{\natexlab{b}})}]{Agnes:2017cz}
\bibinfo{author}{\bibfnamefont{P.}~\bibnamefont{Agnes}}  \bibnamefont{et~al.}
  (\bibinfo{affiliation}{The DarkSide Collaboration}),
  \href{http://dx.doi.org/10.1088/1748-0221/12/10/P10015}{\bibinfo{journal}{JINST}
  \textbf{\bibinfo{volume}{12}}, \bibinfo{pages}{P10015}\bibinfo{year}{
  (\bibinfo{year}{2017}{\natexlab{b}})}}.

\bibitem[{\citenamefont{Agostinelli et~al.}(2003)}]{Agostinelli:2003fg}
\bibinfo{author}{\bibfnamefont{S.}~\bibnamefont{Agostinelli}}
  \bibnamefont{et~al.},
  \href{http://dx.doi.org/10.1016/S0168-9002(03)01368-8}{\bibinfo{journal}{Nucl.
  Inst. Meth. A} \textbf{\bibinfo{volume}{506}},
  \bibinfo{pages}{250}\bibinfo{year}{ (\bibinfo{year}{2003})}}.

\bibitem[{\citenamefont{Allison et~al.}(2006)}]{Allison:2006cd}
\bibinfo{author}{\bibfnamefont{J.}~\bibnamefont{Allison}}
  \bibnamefont{et~al.},
  \href{http://dx.doi.org/10.1109/TNS.2006.869826}{\bibinfo{journal}{IEEE
  Trans. Nucl. Sci.} \textbf{\bibinfo{volume}{53}},
  \bibinfo{pages}{270}\bibinfo{year}{ (\bibinfo{year}{2006})}}.

\bibitem[{\citenamefont{Agnes et~al.}(2017{\natexlab{c}})}]{Agnes:2017ck}
\bibinfo{author}{\bibfnamefont{P.}~\bibnamefont{Agnes}}  \bibnamefont{et~al.}
  (\bibinfo{affiliation}{The DarkSide Collaboration}),
  \href{http://dx.doi.org/10.1088/1748-0221/12/12/P12011}{\bibinfo{journal}{JINST}
  \textbf{\bibinfo{volume}{12}}, \bibinfo{pages}{P12011}\bibinfo{year}{
  (\bibinfo{year}{2017}{\natexlab{c}})}}.

\bibitem[{\citenamefont{Fan}(2016)}]{Fan:2016ud}
\bibinfo{author}{\bibfnamefont{A.}~\bibnamefont{Fan}}
  (\bibinfo{affiliation}{University of California at Los Angeles}), Ph.D.
  thesis, \bibinfo{school}{University of California at Los Angeles}
  (\bibinfo{year}{2016}),
  \urlprefix\url{https://search.proquest.com/docview/1823196595}.

\bibitem[{\citenamefont{Agnes et~al.}(2018{\natexlab{a}})}]{Agnes:2018vt}
\bibinfo{author}{\bibfnamefont{P.}~\bibnamefont{Agnes}}  \bibnamefont{et~al.}
  (\bibinfo{affiliation}{The DarkSide Collaboration}),
  \href{http://arxiv.org/abs/1802.01427v1}{\bibinfo{journal}{arXiv}:1802.01427v1\bibinfo{year}{
  (\bibinfo{year}{2018}{\natexlab{a}})}}.

\bibitem[{\citenamefont{Aprile et~al.}(2014)}]{Aprile:2014he}
\bibinfo{author}{\bibfnamefont{E.}~\bibnamefont{Aprile}}  \bibnamefont{et~al.},
   \href{http://dx.doi.org/10.1088/0954-3899/41/3/035201}{\bibinfo{journal}{J.
  Phys. G} \textbf{\bibinfo{volume}{41}},
  \bibinfo{pages}{035201}\bibinfo{year}{ (\bibinfo{year}{2014})}}.

\bibitem[{\citenamefont{Sorensen}(2017)}]{Sorensen:2017wo}
\bibinfo{author}{\bibfnamefont{P.}~\bibnamefont{Sorensen}},
  \href{http://arxiv.org/abs/1702.04805v1}{\bibinfo{journal}{arXiv}:1702.04805v1\bibinfo{year}{
  (\bibinfo{year}{2017})}}.

\bibitem[{\citenamefont{Sorensen and Kamdin}(2017)}]{Sorensen:2017tf}
\bibinfo{author}{\bibfnamefont{P.}~\bibnamefont{Sorensen}} \bibnamefont{and}
  \bibinfo{author}{\bibfnamefont{K.}~\bibnamefont{Kamdin}},
  \href{http://arxiv.org/abs/1711.07025v2}{\bibinfo{journal}{arXiv}:1711.07025v2\bibinfo{year}{
  (\bibinfo{year}{2017})}}.

\bibitem[{\citenamefont{Kirkwood et~al.}(1948)\citenamefont{Kirkwood,
  Pontecorvo, and Hanna}}]{Kirkwood:1948fg}
\bibinfo{author}{\bibfnamefont{D.~H.~W.} \bibnamefont{Kirkwood}},
  \bibinfo{author}{\bibfnamefont{B.}~\bibnamefont{Pontecorvo}},
  \bibnamefont{and} \bibinfo{author}{\bibfnamefont{G.~C.} \bibnamefont{Hanna}},
   \href{http://dx.doi.org/10.1103/PhysRev.74.497}{\bibinfo{journal}{Phys.
  Rev.} \textbf{\bibinfo{volume}{74}}, \bibinfo{pages}{497}\bibinfo{year}{
  (\bibinfo{year}{1948})}}.

\bibitem[{\citenamefont{Pontecorvo et~al.}(1949)\citenamefont{Pontecorvo,
  Kirkwood, and Hanna}}]{Pontecorvo:1949kb}
\bibinfo{author}{\bibfnamefont{B.}~\bibnamefont{Pontecorvo}},
  \bibinfo{author}{\bibfnamefont{D.~H.~W.} \bibnamefont{Kirkwood}},
  \bibnamefont{and} \bibinfo{author}{\bibfnamefont{G.~C.} \bibnamefont{Hanna}},
   \href{http://dx.doi.org/10.1103/PhysRev.75.982}{\bibinfo{journal}{Phys.
  Rev.} \textbf{\bibinfo{volume}{75}}, \bibinfo{pages}{982}\bibinfo{year}{
  (\bibinfo{year}{1949})}}.

\bibitem[{\citenamefont{Hanna et~al.}(1949)\citenamefont{Hanna, Kirkwood, and
  Pontecorvo}}]{Hanna:1949iv}
\bibinfo{author}{\bibfnamefont{G.~C.} \bibnamefont{Hanna}},
  \bibinfo{author}{\bibfnamefont{D.~H.~W.} \bibnamefont{Kirkwood}},
  \bibnamefont{and}
  \bibinfo{author}{\bibfnamefont{B.}~\bibnamefont{Pontecorvo}},
  \href{http://dx.doi.org/10.1103/PhysRev.75.985.2}{\bibinfo{journal}{Phys.
  Rev.} \textbf{\bibinfo{volume}{75}}, \bibinfo{pages}{985}\bibinfo{year}{
  (\bibinfo{year}{1949})}}.

\bibitem[{\citenamefont{Firestone et~al.}(1999)\citenamefont{Firestone, Baglin,
  and Chu}}]{Firestone:1999tz}
\bibinfo{author}{\bibfnamefont{R.~B.} \bibnamefont{Firestone}},
  \bibinfo{author}{\bibfnamefont{C.~M.} \bibnamefont{Baglin}},
  \bibnamefont{and} \bibinfo{author}{\bibfnamefont{S.~Y.~F.}
  \bibnamefont{Chu}},
  \href{http://books.google.com/books?id=W4JUAAAAMAAJ&q=firestone+table+of+isotopes&dq=firestone+table+of+isotopes&hl=&cd=1&source=gbs_api}{\emph{\bibinfo{title}{{Table
  of
  isotopes}}}}\href{http://books.google.com/books?id=W4JUAAAAMAAJ&q=firestone+table+of+isotopes&dq=firestone+table+of+isotopes&hl=&cd=1&source=gbs_api}{,
  \bibinfo{publisher}{Wiley-Interscience} (\bibinfo{year}{1999})}.

\bibitem[{\citenamefont{Odiot and Daudel}(1956)}]{Odiot:1956ha}
\bibinfo{author}{\bibfnamefont{S.}~\bibnamefont{Odiot}} \bibnamefont{and}
  \bibinfo{author}{\bibfnamefont{R.}~\bibnamefont{Daudel}},
  \href{http://dx.doi.org/10.1051/jphysrad:0195600170106000}{\bibinfo{journal}{J.
  Phys. Radium} \textbf{\bibinfo{volume}{17}},
  \bibinfo{pages}{60}\bibinfo{year}{ (\bibinfo{year}{1956})}}.

\bibitem[{\citenamefont{Brysk and Rose}(1958)}]{Brysk:1958ky}
\bibinfo{author}{\bibfnamefont{H.}~\bibnamefont{Brysk}} \bibnamefont{and}
  \bibinfo{author}{\bibfnamefont{M.~E.} \bibnamefont{Rose}},
  \href{http://dx.doi.org/10.1103/RevModPhys.30.1169}{\bibinfo{journal}{Rev.
  Mod. Phys.} \textbf{\bibinfo{volume}{30}},
  \bibinfo{pages}{1169}\bibinfo{year}{ (\bibinfo{year}{1958})}}.

\bibitem[{\citenamefont{Santos-Ocampo and Conway}(1960)}]{SantosOcampo:1960jaa}
\bibinfo{author}{\bibfnamefont{A.~G.} \bibnamefont{Santos-Ocampo}}
  \bibnamefont{and} \bibinfo{author}{\bibfnamefont{D.~C.}
  \bibnamefont{Conway}},
  \href{http://dx.doi.org/10.1103/PhysRev.120.2196}{\bibinfo{journal}{Phys.
  Rev.} \textbf{\bibinfo{volume}{120}}, \bibinfo{pages}{2196}\bibinfo{year}{
  (\bibinfo{year}{1960})}}.

\bibitem[{\citenamefont{Sangiorgio et~al.}(2013)}]{Sangiorgio:2013gh}
\bibinfo{author}{\bibfnamefont{S.}~\bibnamefont{Sangiorgio}}
  \bibnamefont{et~al.},
  \href{http://dx.doi.org/10.1016/j.nima.2013.06.061}{\bibinfo{journal}{Nucl.
  Inst. Meth. A} \textbf{\bibinfo{volume}{728}},
  \bibinfo{pages}{69}\bibinfo{year}{ (\bibinfo{year}{2013})}}.

\bibitem[{\citenamefont{Edkins~Ludert}(2017)}]{Edkins:2017vf}
\bibinfo{author}{\bibfnamefont{E.}~\bibnamefont{Edkins~Ludert}}
  (\bibinfo{affiliation}{University of Hawai'i at Manoa}), Ph.D. thesis,
  \bibinfo{school}{University of Hawai'i at Manoa} (\bibinfo{year}{2017}),
  \urlprefix\url{https://search.proquest.com/docview/1953252158/}.

\bibitem[{\citenamefont{Alexander et~al.}(2013)}]{Alexander:2013ke}
\bibinfo{author}{\bibfnamefont{T.}~\bibnamefont{Alexander}}
  \bibnamefont{et~al.} (\bibinfo{affiliation}{The SCENE Collaboration}),
  \href{http://dx.doi.org/10.1103/PhysRevD.88.092006}{\bibinfo{journal}{Phys.
  Rev. D} \textbf{\bibinfo{volume}{88}}, \bibinfo{pages}{092006}\bibinfo{year}{
  (\bibinfo{year}{2013})}}.

\bibitem[{\citenamefont{Cao et~al.}(2015)}]{Cao:2015ks}
\bibinfo{author}{\bibfnamefont{H.}~\bibnamefont{Cao}}  \bibnamefont{et~al.}
  (\bibinfo{affiliation}{The SCENE Collaboration}),
  \href{http://dx.doi.org/10.1103/PhysRevD.91.092007}{\bibinfo{journal}{Phys.
  Rev. D} \textbf{\bibinfo{volume}{91}}, \bibinfo{pages}{092007}\bibinfo{year}{
  (\bibinfo{year}{2015})}}.

\bibitem[{\citenamefont{Agnes et~al.}(2018{\natexlab{b}})}]{Agnes:2018uo}
\bibinfo{author}{\bibfnamefont{P.}~\bibnamefont{Agnes}}  \bibnamefont{et~al.}
  (\bibinfo{affiliation}{The ARIS Collaboration}),
  \href{http://arxiv.org/abs/1801.06653v1}{\bibinfo{journal}{arXiv}:1801.06653v1\bibinfo{year}{
  (\bibinfo{year}{2018}{\natexlab{b}})}}.

\bibitem[{\citenamefont{Bezrukov et~al.}(2011)\citenamefont{Bezrukov,
  Kahlhoefer, and Lindner}}]{Bezrukov:2011gi}
\bibinfo{author}{\bibfnamefont{F.}~\bibnamefont{Bezrukov}},
  \bibinfo{author}{\bibfnamefont{F.}~\bibnamefont{Kahlhoefer}},
  \bibnamefont{and} \bibinfo{author}{\bibfnamefont{M.}~\bibnamefont{Lindner}},
  \href{http://dx.doi.org/10.1016/j.astropartphys.2011.06.008}{\bibinfo{journal}{Astropart.
  Phys.} \textbf{\bibinfo{volume}{35}}, \bibinfo{pages}{119}\bibinfo{year}{
  (\bibinfo{year}{2011})}}.

\bibitem[{\citenamefont{Lindhard et~al.}(1963)\citenamefont{Lindhard, Scharff,
  and Schi{\o}tt}}]{Lindhard:1963tr}
\bibinfo{author}{\bibfnamefont{J.}~\bibnamefont{Lindhard}},
  \bibinfo{author}{\bibfnamefont{M.}~\bibnamefont{Scharff}},  \bibnamefont{and}
  \bibinfo{author}{\bibfnamefont{H.~E.} \bibnamefont{Schi{\o}tt}},
  \href{http://gymarkiv.sdu.dk/MFM/meddelelser.htm}{\bibinfo{journal}{Det Kgl.
  Danske Viden.} \textbf{\bibinfo{volume}{33}},
  \bibinfo{pages}{14:1}\bibinfo{year}{ (\bibinfo{year}{1963})}}.

\bibitem[{\citenamefont{Joshi et~al.}(2014)}]{Joshi:2014kt}
\bibinfo{author}{\bibfnamefont{T.~H.} \bibnamefont{Joshi}}
  \bibnamefont{et~al.},
  \href{http://dx.doi.org/10.1103/PhysRevLett.112.171303}{\bibinfo{journal}{Phys.
  Rev. Lett.} \textbf{\bibinfo{volume}{112}},
  \bibinfo{pages}{171303}\bibinfo{year}{ (\bibinfo{year}{2014})}}.

\bibitem[{\citenamefont{{After consulting with the authors, the data point is
  corrected for their single electron yield using the 2.82 keV K-shell capture
  \ce{^37Ar} line from their experiment and DarkSide-50 as a cross-calibration
  point}}(Feb. 2018)}]{Joshi:2018pc}
\bibinfo{author}{\bibnamefont{{After consulting with the authors, the data
  point is corrected for their single electron yield using the 2.82 keV K-shell
  capture \ce{^37Ar} line from their experiment and DarkSide-50 as a
  cross-calibration point}}}, \bibinfo{howpublished}{private communication with
  T. H. Joshi and S. Sangiorgio} (\bibinfo{year}{Feb. 2018}).

\bibitem[{\citenamefont{Smith et~al.}(2007)}]{Smith:2007fi}
\bibinfo{author}{\bibfnamefont{M.~C.} \bibnamefont{Smith}}
  \bibnamefont{et~al.},
  \href{http://dx.doi.org/10.1111/j.1365-2966.2007.11964.x}{\bibinfo{journal}{Month.
  Not. Royal Astron. Soc.} \textbf{\bibinfo{volume}{379}},
  \bibinfo{pages}{755}\bibinfo{year}{ (\bibinfo{year}{2007})}}.

\bibitem[{\citenamefont{Cowan et~al.}(2011)\citenamefont{Cowan, Cranmer, Gross,
  and Vitells}}]{Cowan:2011cx}
\bibinfo{author}{\bibfnamefont{G.}~\bibnamefont{Cowan}},
  \bibinfo{author}{\bibfnamefont{K.}~\bibnamefont{Cranmer}},
  \bibinfo{author}{\bibfnamefont{E.}~\bibnamefont{Gross}},  \bibnamefont{and}
  \bibinfo{author}{\bibfnamefont{O.}~\bibnamefont{Vitells}},
  \href{http://dx.doi.org/10.1140/epjc/s10052-011-1554-0}{\bibinfo{journal}{Eur.
  Phys. J. C} \textbf{\bibinfo{volume}{71}}, \bibinfo{pages}{1}\bibinfo{year}{
  (\bibinfo{year}{2011})}}.

\bibitem[{\citenamefont{Moneta et~al.}(2011)\citenamefont{Moneta, Cranmer,
  Schott, and Ververke}}]{Moneta:2011jc}
\bibinfo{author}{\bibfnamefont{L.}~\bibnamefont{Moneta}},
  \bibinfo{author}{\bibfnamefont{K.}~\bibnamefont{Cranmer}},
  \bibinfo{author}{\bibfnamefont{G.}~\bibnamefont{Schott}},  \bibnamefont{and}
  \bibinfo{author}{\bibfnamefont{W.}~\bibnamefont{Ververke}},
  \href{http://dx.doi.org/10.22323/1.093.0057}{\bibinfo{journal}{Proc. Sci.}
  \textbf{\bibinfo{volume}{93}}, \bibinfo{pages}{057}\bibinfo{year}{
  (\bibinfo{year}{2011})}}.

\bibitem[{\citenamefont{Verkerke and Kirkby}(2006)}]{Verkerke:2006es}
\bibinfo{author}{\bibfnamefont{W.}~\bibnamefont{Verkerke}} \bibnamefont{and}
  \bibinfo{author}{\bibfnamefont{D.}~\bibnamefont{Kirkby}},
  \href{http://dx.doi.org/10.1142/9781860948985_0039}{\bibinfo{journal}{Proc.
  PHYSTAT05} \textbf{\bibinfo{volume}{1}}, \bibinfo{pages}{186}\bibinfo{year}{
  (\bibinfo{year}{2006})}}.

\bibitem[{\citenamefont{Agnese et~al.}(2014{\natexlab{a}})}]{Agnese:2014cq}
\bibinfo{author}{\bibfnamefont{R.}~\bibnamefont{Agnese}}  \bibnamefont{et~al.}
  (\bibinfo{affiliation}{The SuperCDMS Collaboration}),
  \href{http://dx.doi.org/10.1103/PhysRevLett.112.041302}{\bibinfo{journal}{Phys.
  Rev. Lett.} \textbf{\bibinfo{volume}{112}},
  \bibinfo{pages}{041302}\bibinfo{year}{
  (\bibinfo{year}{2014}{\natexlab{a}})}}.

\bibitem[{\citenamefont{Agnese et~al.}(2014{\natexlab{b}})}]{Agnese:2014ix}
\bibinfo{author}{\bibfnamefont{R.}~\bibnamefont{Agnese}}  \bibnamefont{et~al.}
  (\bibinfo{affiliation}{The SuperCDMS Collaboration}),
  \href{http://dx.doi.org/10.1103/PhysRevLett.112.241302}{\bibinfo{journal}{Phys.
  Rev. Lett.} \textbf{\bibinfo{volume}{112}},
  \bibinfo{pages}{241302}\bibinfo{year}{
  (\bibinfo{year}{2014}{\natexlab{b}})}}.

\bibitem[{\citenamefont{Angloher et~al.}(2016)}]{Angloher:2016js}
\bibinfo{author}{\bibfnamefont{G.}~\bibnamefont{Angloher}}
  \bibnamefont{et~al.} (\bibinfo{affiliation}{The CRESST Collaboration}),
  \href{http://dx.doi.org/10.1140/epjc/s10052-016-3877-3}{\bibinfo{journal}{Eur.
  Phys. J. C} \textbf{\bibinfo{volume}{76}}, \bibinfo{pages}{25}\bibinfo{year}{
  (\bibinfo{year}{2016})}}.

\bibitem[{\citenamefont{Zhao et~al.}(2016)}]{Zhao:2016bq}
\bibinfo{author}{\bibfnamefont{W.}~\bibnamefont{Zhao}}  \bibnamefont{et~al.}
  (\bibinfo{affiliation}{The CDEX Collaboration}),
  \href{http://dx.doi.org/10.1103/PhysRevD.93.092003}{\bibinfo{journal}{Phys.
  Rev. D} \textbf{\bibinfo{volume}{93}}, \bibinfo{pages}{092003}\bibinfo{year}{
  (\bibinfo{year}{2016})}}.

\bibitem[{\citenamefont{Hehn et~al.}(2016)}]{Hehn:2016eq}
\bibinfo{author}{\bibfnamefont{L.}~\bibnamefont{Hehn}}  \bibnamefont{et~al.}
  (\bibinfo{affiliation}{The EDELWEISS Collaboration}),
  \href{http://dx.doi.org/10.1140/epjc/s10052-016-4388-y}{\bibinfo{journal}{Eur.
  Phys. J. C} \textbf{\bibinfo{volume}{76}},
  \bibinfo{pages}{548}\bibinfo{year}{ (\bibinfo{year}{2016})}}.

\bibitem[{\citenamefont{Aguilar-Arevalo et~al.}(2016)}]{AguilarArevalo:2016ei}
\bibinfo{author}{\bibfnamefont{A.}~\bibnamefont{Aguilar-Arevalo}}
  \bibnamefont{et~al.} (\bibinfo{affiliation}{The DAMIC Collaboration}),
  \href{http://dx.doi.org/10.1103/PhysRevD.94.082006}{\bibinfo{journal}{Phys.
  Rev. D} \textbf{\bibinfo{volume}{94}}, \bibinfo{pages}{082006}\bibinfo{year}{
  (\bibinfo{year}{2016})}}.

\bibitem[{\citenamefont{Tan et~al.}(2016)}]{Tan:2016fv}
\bibinfo{author}{\bibfnamefont{A.}~\bibnamefont{Tan}}  \bibnamefont{et~al.}
  (\bibinfo{affiliation}{The PandaX-II Collaboration}),
  \href{http://dx.doi.org/10.1103/PhysRevLett.117.121303}{\bibinfo{journal}{Phys.
  Rev. Lett.} \textbf{\bibinfo{volume}{117}},
  \bibinfo{pages}{121303}\bibinfo{year}{ (\bibinfo{year}{2016})}}.

\bibitem[{\citenamefont{Petricca et~al.}(2017)}]{Petricca:2017vx}
\bibinfo{author}{\bibfnamefont{F.}~\bibnamefont{Petricca}}
  \bibnamefont{et~al.} (\bibinfo{affiliation}{The CRESST Collaboration}),
  \href{http://arxiv.org/abs/1711.07692v1}{\bibinfo{journal}{arXiv}:1711.07692v1\bibinfo{year}{
  (\bibinfo{year}{2017})}}.

\bibitem[{\citenamefont{Agnese et~al.}(2017)}]{Agnese:2017wy}
\bibinfo{author}{\bibfnamefont{R.}~\bibnamefont{Agnese}}  \bibnamefont{et~al.}
  (\bibinfo{affiliation}{The SuperCDMS Collaboration}),
  \href{http://arxiv.org/abs/1707.01632v2}{\bibinfo{journal}{arXiv}:1707.01632v2\bibinfo{year}{
  (\bibinfo{year}{2017})}}.

\bibitem[{\citenamefont{Behnke et~al.}(2017)}]{Behnke:2017cc}
\bibinfo{author}{\bibfnamefont{E.}~\bibnamefont{Behnke}}  \bibnamefont{et~al.}
  (\bibinfo{affiliation}{The PICASSO Collaboration}),
  \href{http://dx.doi.org/10.1016/j.astropartphys.2017.02.005}{\bibinfo{journal}{Astropart.
  Phys.} \textbf{\bibinfo{volume}{90}}, \bibinfo{pages}{85}\bibinfo{year}{
  (\bibinfo{year}{2017})}}.

\bibitem[{\citenamefont{Amole et~al.}(2017)}]{Amole:2017iz}
\bibinfo{author}{\bibfnamefont{C.}~\bibnamefont{Amole}}  \bibnamefont{et~al.}
  (\bibinfo{affiliation}{The PICO Collaboration}),
  \href{http://dx.doi.org/10.1103/PhysRevLett.118.251301}{\bibinfo{journal}{Phys.
  Rev. Lett.} \textbf{\bibinfo{volume}{118}},
  \bibinfo{pages}{251301}\bibinfo{year}{ (\bibinfo{year}{2017})}}.

\bibitem[{\citenamefont{Aprile et~al.}(2017)}]{Aprile:2017eq}
\bibinfo{author}{\bibfnamefont{E.}~\bibnamefont{Aprile}}  \bibnamefont{et~al.}
  (\bibinfo{affiliation}{The XENON Collaboration}),
  \href{http://dx.doi.org/10.1103/PhysRevLett.119.181301}{\bibinfo{journal}{Phys.
  Rev. Lett.} \textbf{\bibinfo{volume}{119}},
  \bibinfo{pages}{181301}\bibinfo{year}{ (\bibinfo{year}{2017})}}.

\bibitem[{\citenamefont{Akerib et~al.}(2017)}]{Akerib:2017kg}
\bibinfo{author}{\bibfnamefont{D.~S.} \bibnamefont{Akerib}}
  \bibnamefont{et~al.} (\bibinfo{affiliation}{The LUX Collaboration}),
  \href{http://dx.doi.org/10.1103/PhysRevLett.118.021303}{\bibinfo{journal}{Phys.
  Rev. Lett.} \textbf{\bibinfo{volume}{118}},
  \bibinfo{pages}{021303}\bibinfo{year}{ (\bibinfo{year}{2017})}}.

\bibitem[{\citenamefont{Arnaud et~al.}(2018)}]{Arnaud:2018fg}
\bibinfo{author}{\bibfnamefont{Q.}~\bibnamefont{Arnaud}}  \bibnamefont{et~al.}
  (\bibinfo{affiliation}{The NEWS-G Collaboration}),
  \href{http://dx.doi.org/10.1016/j.astropartphys.2017.10.009}{\bibinfo{journal}{Astropart.
  Phys.} \textbf{\bibinfo{volume}{97}}, \bibinfo{pages}{54}\bibinfo{year}{
  (\bibinfo{year}{2018})}}.

\bibitem[{\citenamefont{Jiang et~al.}(2018)}]{Jiang:2018ez}
\bibinfo{author}{\bibfnamefont{H.}~\bibnamefont{Jiang}}  \bibnamefont{et~al.}
  (\bibinfo{affiliation}{The CDEX Collaboration}),
  \href{http://dx.doi.org/10.1103/PhysRevLett.120.241301}{\bibinfo{journal}{Phys.
  Rev. Lett.} \textbf{\bibinfo{volume}{120}},
  \bibinfo{pages}{241301}\bibinfo{year}{ (\bibinfo{year}{2018})}}.

\bibitem[{\citenamefont{Savage et~al.}(2009)\citenamefont{Savage, Gelmini,
  Gondolo, and Freese}}]{Savage:2009hi}
\bibinfo{author}{\bibfnamefont{C.}~\bibnamefont{Savage}},
  \bibinfo{author}{\bibfnamefont{G.}~\bibnamefont{Gelmini}},
  \bibinfo{author}{\bibfnamefont{P.}~\bibnamefont{Gondolo}},  \bibnamefont{and}
  \bibinfo{author}{\bibfnamefont{K.}~\bibnamefont{Freese}},
  \href{http://dx.doi.org/10.1088/1475-7516/2009/04/010}{\bibinfo{journal}{JCAP}
  \textbf{\bibinfo{volume}{2009}}, \bibinfo{pages}{010}\bibinfo{year}{
  (\bibinfo{year}{2009})}}.

\bibitem[{\citenamefont{Angloher et~al.}(2012)}]{Angloher:2012kl}
\bibinfo{author}{\bibfnamefont{G.}~\bibnamefont{Angloher}}
  \bibnamefont{et~al.} (\bibinfo{affiliation}{The CRESST Collaboration}),
  \href{http://dx.doi.org/10.1140/epjc/s10052-012-1971-8}{\bibinfo{journal}{Eur.
  Phys. J. C} \textbf{\bibinfo{volume}{72}},
  \bibinfo{pages}{1971}\bibinfo{year}{ (\bibinfo{year}{2012})}}.

\bibitem[{\citenamefont{Aalseth et~al.}(2013)}]{Aalseth:2013bg}
\bibinfo{author}{\bibfnamefont{C.~E.} \bibnamefont{Aalseth}}
  \bibnamefont{et~al.} (\bibinfo{affiliation}{The CoGeNT Collaboration}),
  \href{http://dx.doi.org/10.1103/PhysRevD.88.012002}{\bibinfo{journal}{Phys.
  Rev. D} \textbf{\bibinfo{volume}{88}}, \bibinfo{pages}{012002}\bibinfo{year}{
  (\bibinfo{year}{2013})}}.

\bibitem[{\citenamefont{Ruppin et~al.}(2014)\citenamefont{Ruppin, Billard,
  Figueroa-Feliciano, and Strigari}}]{Ruppin:2014ee}
\bibinfo{author}{\bibfnamefont{F.}~\bibnamefont{Ruppin}},
  \bibinfo{author}{\bibfnamefont{J.}~\bibnamefont{Billard}},
  \bibinfo{author}{\bibfnamefont{E.}~\bibnamefont{Figueroa-Feliciano}},
  \bibnamefont{and} \bibinfo{author}{\bibfnamefont{L.}~\bibnamefont{Strigari}},
   \href{http://dx.doi.org/10.1103/PhysRevD.90.083510}{\bibinfo{journal}{Phys.
  Rev. D} \textbf{\bibinfo{volume}{90}}, \bibinfo{pages}{083510}\bibinfo{year}{
  (\bibinfo{year}{2014})}}.

\end{thebibliography}
\end{document}